\documentclass[fleqn,usenatbib]{mnras}

\usepackage{newtxtext,newtxmath}
 
\usepackage[T1]{fontenc}

\DeclareRobustCommand{\VAN}[3]{#2}
\let\VANthebibliography\thebibliography
\def\thebibliography{\DeclareRobustCommand{\VAN}[3]{##3}\VANthebibliography}

\usepackage{graphicx}	% Including figure files
\usepackage{amsmath}	% Advanced maths commands
\usepackage{amssymb}	% Extra maths symbols
\usepackage{xfrac}      % Works better with other fonts

\graphicspath{ {./figures/} }

\def\adaptahop    {{\sc AdaptaHop}}
\def\bpass    {{\sc Bpass}}
\def\cloudy    {{\sc Cloudy}}
\def\ramses    {{\sc Ramses}}
\def\sphinx  {{\sc Sphinx}}
\def\cci {{\rm{cm}}^{-3}}   % cm-3
\def\mdm {m_{\rm DM}}
\def\msun {M_{\odot}}
\def\nh {n_{\rm{H}}}

\title[The Biermann Battery during the EoR]{Cosmological Magnetogenesis: The Biermann Battery during the Epoch of Reionization}

\author[M. Attia et al.]{
Mara Attia,$^{1,2}$\thanks{E-mail: mara.attia@unige.ch}
Romain Teyssier,$^2$
Harley Katz,$^3$
Taysun Kimm,$^4$
Sergio Martin-Alvarez,$^5$
\newauthor{}
Pierre Ocvirk,$^6$
and Joakim Rosdahl$^7$
\\
$^1$Department of Physics, ETH Zurich, Wolfgang-Pauli-Strasse 27, CH-8093 Zurich, Switzerland\\
$^2$Institute for Computational Science, University of Zurich, Winterthurerstrasse 190, CH-8057 Zurich, Switzerland\\
$^3$Sub-department of Astrophysics, University of Oxford, Keble Road, Oxford OX1 3RH, UK\\
$^4$Department of Astronomy, Yonsei University, 50 Yonsei-ro, Seodaemun-gu, Seoul 03722, Republic of Korea\\
$^5$Kavli Institute for Cosmology and Institute of Astronomy, Madingley Road, Cambridge CB3 0HA, UK\\
$^6$Universit\'e de Strasbourg, CNRS, Observatoire astronomique de Strasbourg (ObAS), UMR 7550, F-67000 Strasbourg, France\\
$^7$Universit\'e Lyon, Univ Lyon1, ENS de Lyon, CNRS, Centre de Recherche Astrophysique de Lyon UMR5574, F-69230 Saint-Genis-Laval, France
}

\date{Accepted XXX. Received YYY; in original form ZZZ}

\pubyear{2021}

\begin{document}
\label{firstpage}
\pagerange{\pageref{firstpage}--\pageref{lastpage}}
\maketitle

\begin{abstract}
We investigate the effect of the Biermann battery during the Epoch of Reionization (EoR) using cosmological Adaptive Mesh Refinement simulations within the framework of the \sphinx{} project. We develop a novel numerical technique to solve for the Biermann battery term in the Constrained Transport method, preserving both the zero divergence of the magnetic field and the absence of Biermann battery for isothermal flows. The structure-preserving nature of our numerical method turns out to be very important to minimise numerical errors during validation tests of the propagation of a Strömgren sphere and of a Sedov blast wave. We then use this new method to model the evolution of a 2.5 and 5 co-moving Mpc cosmological box with a state-of-the-art galaxy formation model within the \ramses{} code. Contrary to previous findings, we show that three different Biermann battery channels emerge: the first one is associated with linear perturbations before the EoR, the second one is the classical Biermann battery associated with reionization fronts during the EoR, and the third one is associated with strong, supernova-driven outflows. While the two former channels generate spontaneously volume-filling magnetic fields with a strength on the order or below $10^{-20}$~G, the latter, owing to the higher plasma temperature and a marginally-resolved turbulent dynamo, reaches a field strength as high as $10^{-18}$~G in the intergalactic medium around massive haloes.
\end{abstract}


%%%%%%%%%%%%%%%%%%%%%%%%%%%%%%%%%%%%%%%%%%%%%%%%%%

%%%%%%%%%%%%%%%%% BODY OF PAPER %%%%%%%%%%%%%%%%%%

\section{Introduction}

Magnetic fields are an important component of the conductive plasma found in many astrophysical objects, such as planets, stars, galaxies, but also in the intergalactic medium (IGM) between galaxies. Their exact origin is however still debated: they could have formed in the early Universe \citep{2013A&ARv..21...62D} or at a later epoch, during structure formation through microscopic processes such as the Biermann battery \citep{Biermann1950} or the Weibel instability \citep{2009JPlPh..75...19L}. Magnetic fields could also be generated inside stars through stellar dynamos or within galaxies through galactic dynamos \citep{2005PhR...417....1B} and then released into the interstellar medium (ISM) by supernova (SN) explosions \citep{1973SvA....17..137B} or by active galactic nuclei (AGN) jets \citep{2005LNP...664....1R}, and then released into the IGM by galactic winds \citep{Bertone2006}.

For the stellar or galactic dynamos to operate, a small seed magnetic field is required. Among many alternative scenarios for such seed fields, the Biermann battery, also known as the `thermo-electric effect', is probably the most robust and therefore the most popular theory. It has been observed in laboratory experiments \citep{2010HEDP....6..162F} and it has been simulated in the cosmological context within accretion shocks \citep{Kulsrud1997} or ionization fronts \citep{Gnedin2000} with fields generated by thermally-driven currents at the level of $10^{-20}$~G.

Galaxy formation models have evolved quite substantially over the last two decades, with the increasing role of SN and AGN feedback processes in regulating star formation and determining the morphological type of galaxies \citep[see][and references therein]{2017ARA&A..55...59N}. Compared to the earlier galaxy formation models, stronger galactic outflows with correspondingly more intense ionizing radiation fields could quantitatively change the conclusions of these earlier works. It has been also pointed out by \cite{Naoz2013} that the Biermann battery could be already effective during the linear evolution of the coupled dark matter (DM)/gas fluid during the dark ages, leading to small seed fields as large as $10^{-24}$~G, even before any star could have formed. 

Moreover, \cite{Graziani2015} recently showed that numerically modelling the Biermann battery can be particularly challenging, and can lead, in the case of sharp discontinuities in the flow, to a spurious numerical effect that they called `the Biermann catastrophe'. These various considerations have led us to revisit the early work of \cite{Kulsrud1997} and \cite{Gnedin2000}, and study the Biermann battery during the Epoch of Reionization (EoR). 

The process of reionizing the Universe after recombination likely began after $z=30$ with the formation of the first stars \citep{2012ApJ...745...50W} and ended in the redshift range $z \sim 5-6$ \citep{2006ARA&A..44..415F}. During this first billion years, the Universe may have progressively reionized due to the growing number of small star-forming galaxies \citep[e.g.][]{2019ApJ...879...36F}. 

A key aspect of numerical models of cosmic reionization is to be able to correctly capture the physical state of the ISM in these dwarf galaxies, so as to resolve the ionized channels through which the photons escape \citep{2014ApJ...788..121K,Kimm2017}. This results in a severe resolution requirement for cosmological simulations that must also feature a coupled treatment of self-gravity, hydrodynamics and radiative transfer. This is precisely the challenges that the \sphinx{} project is addressing \citep{Rosdahl2018}. The goal of this paper is to model the Biermann battery process and to study the corresponding generated magnetic fields, using the computational infrastructure of the \sphinx{} project. This work is complementary to the \textsc{Sphinx-MHD} study \citep{Katz2021}, which focuses on primordial magnetic fields during the EoR and their impact on galaxy formation and cosmic reionization.

In this paper, we first describe in Sect. \ref{sect:methods} a new numerical method designed to specifically address the numerical challenges associated with the Biermann battery process. In Sect. \ref{sect:tests}, we perform static tests to ensure the correct implementation of the module, as well as detailed tests of the propagation of a Strömgren sphere, featuring sharp ionization fronts, and a Sedov blast wave with large temperature discontinuities. We compare our new method to the traditional naive implementation and show that these typical flow configurations can indeed be modelled accurately without suffering from the Biermann catastrophe. We also present in Sect. \ref{sect:cosmo} the results of a cosmological simulation of the EoR with the \textsc{Ramses-RT} code \citep{Teyssier2002, Rosdahl2013}, that combines radiative transfer of ionizing radiation to a self-consistent model of galaxy formation with star formation and associated SN feedback. We finally discuss our results in Sect. \ref{sect:discuss}.

%%%%%%%%%%%%%%%%%%%%%%%%%%%%%%%%%%%%%%%%%%%%%%%%%%

\section{Numerical implementation}
\label{sect:methods}

We have implemented a novel numerical scheme for the Biermann battery in the Adaptive Mesh Refinement (AMR) code \ramses{} \citep{Teyssier2002}, which solves the ideal magneto-hydrodynamics (MHD) equations \citep{Teyssier2006,Fromang2006}. The temporal evolution of the magnetic field $\boldsymbol{B}$ is obtained by solving the induction equation
\begin{equation}
\label{eq:induc}
\frac{\partial{\boldsymbol{B}}}{\partial{t}}=\boldsymbol{\nabla}\times\boldsymbol{E},
\end{equation}
where the electromotive force (EMF) is given by $\boldsymbol{E}=\boldsymbol{u}\times\boldsymbol{B}$, with $\boldsymbol{u}$ the gas velocity. Equation (\ref{eq:induc}) is at the core of many important physical processes in cosmic magnetism such as large-scale and small-scale galactic dynamos \citep{2005PhR...417....1B}. In the ideal MHD limit, however, we cannot create magnetic fields in a previously unmagnetised fluid. Our interest precisely being to create seed magnetic fields for subsequent dynamo processes, we need an additional, non-ideal mechanism such as the Biermann battery \citep{Biermann1950}. The corresponding non-ideal source term is written as
\begin{equation}
\label{eq:battery}
\frac{\partial{\boldsymbol{B}}}{\partial{t}}=\boldsymbol{\nabla}\times\boldsymbol{E}_\mathrm{B}\equiv\boldsymbol{\nabla}\times\left(\frac{c}{en_\mathrm{e}}\boldsymbol{\nabla}p_\mathrm{e}\right),
\end{equation}
where $n_{\rm e}$ and $p_{\rm e}$ are respectively the electron number density and the electron pressure, $e$ is the elementary charge of the electron, and $c$ is the speed of light. For the latter, note that we use the actual speed of light, not the reduced speed of light in the radiative transfer module (see Sect. \ref{sect:sphinx}). The Biermann term stems from microscopic electric currents induced by the large difference in the electron and ion mass and is able to generate magnetic fields \textit{ex nihilo} as long as $\boldsymbol{\nabla} n_{\rm e}$ and $\boldsymbol{\nabla}p_{\rm e}$ are misaligned. In the \ramses{} code, we just add the non-ideal Biermann EMF, $\boldsymbol{E}_{\rm B}$, to the ideal EMF, $\boldsymbol{E}$, naturally preserving the divergence-free nature of our scheme.

The \ramses{} code involves a tree-based data structure allowing recursive grid refinements on a cell-by-cell basis. The MHD equations are solved using the second-order unsplit Godunov method based on the monotonic upstream-centred scheme for conservation laws \citep[MUSCL-Hancock method,][]{vanLeer1997,Toro2013}. The Constrained Transport (CT) approach \citep[][]{Yee1966,Evans1988} is used to advance Eq.~(\ref{eq:induc}) in time, which satisfies the solenoidal constraint $\boldsymbol{\nabla}\cdot\boldsymbol{B}=0$ to machine-round-off precision \citep{Teyssier2006}. It simply consists in writing the induction equation in the following conservative integral form on the cell faces:
\begin{equation}
\frac{\partial}{\partial{t}}\iint\boldsymbol{B}\cdot\mathrm{d}\boldsymbol{S}=\oint\boldsymbol{E}\cdot\mathrm{d}\boldsymbol{l},
\end{equation}
where ${\rm d}\boldsymbol{S}$ and ${\rm d}\boldsymbol{l}$ are respectively an elementary surface and length. As a consequence, while all the hydrodynamic variables (density, velocities, total energy) are located at cell centres, this approach requires the magnetic field components to lie on the cell faces. In this framework, the EMF, $\boldsymbol{E}$, is defined at cell edges, so that its circulation defines the $\boldsymbol{B}$ flux across face surfaces. Consequently, the Biermann battery term $\boldsymbol{E}_{\rm B}$ must also be defined at cell edges.

\begin{figure}
\centering
\includegraphics[width=8.5cm]{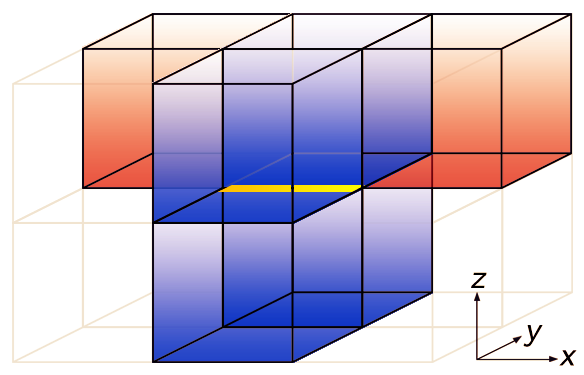}
\caption{Illustration of the naive implementation of the Biermann battery. The orange cells are used to estimate $\partial_x p_{\rm e}$ at the centre of the purple cell they frame. $E_{\mathrm{B,}x}$ is computed at the golden edge by averaging the 4 centred values of the purple cells.}
\label{fig:naive}
\end{figure}

The most straightforward way to proceed is to first calculate the Biermann term in cell centres, where the quantities involved in its definition are located. To this purpose, $\boldsymbol{\nabla} p_{\rm e}$ is computed using the two adjacent cells in the gradient direction. As an example, the expression of the $x$-directed Biermann battery term at the centre of the $(i,j,k)$ cell is
\begin{equation}
E_{\mathrm{B,}x}^{i,j,k}=\frac{c}{en_\mathrm{e}^{i,j,k}}\frac{p_\mathrm{e}^{i+1,j,k}-p_\mathrm{e}^{i-1,j,k}}{2\Delta{}x},
\end{equation} 
where $\Delta x$ is the cell width in the $x$-direction. $\boldsymbol{E}_{\rm B}$ is then derived on each edge by averaging the values at the centre of the 4 adjacent cells (Fig.~\ref{fig:naive}). While this naive implementation has the advantage of being simple, it is none the less vulnerable to spurious numerical effects that emerge at sharp discontinuities in the fluid variables. Indeed, the numerical Biermann magnetic field can diverge with increasing resolution at a discontinuity, resulting into what was called the Biermann catastrophe by \cite{Graziani2015}. In addition, this naive discretisation of the Biermann current can generate spurious fields in regions where $\boldsymbol{\nabla} n_{\rm e}$ and $\boldsymbol{\nabla} p_{\rm e}$ are aligned because of unavoidable truncation errors in the gradient calculations.

\begin{figure}
\centering
\includegraphics[width=8.5cm]{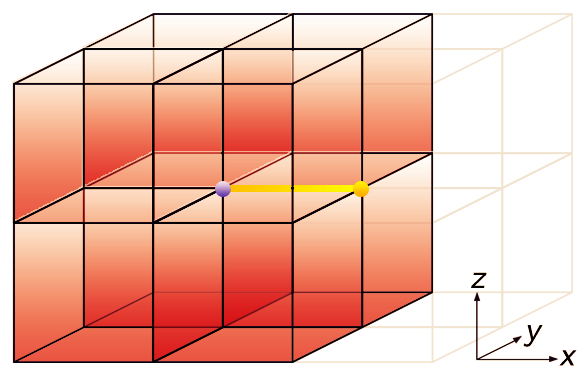}
\caption{Illustration of the improved implementation of the Biermann battery. The orange cells are used to calculate $p_{\rm e}$ and $T_{\rm e}$ at the purple vertex. $E_{\mathrm{B,}x}$ is computed at the golden edge by using the purple and golden end vertices.}
\label{fig:improved}
\end{figure}

The alternative numerical implementation of the Biermann term we propose is based on the following simple manipulation, using the electron fluid equation of state $p_{\rm e}=n_{\rm e}k_{\rm B}T_{\rm e}$,
\begin{equation}
\boldsymbol{E}_\mathrm{B}=\frac{ck_\mathrm{B}}{e}\left(T_\mathrm{e}\boldsymbol{\nabla}\ln{}n_\mathrm{e}+\boldsymbol{\nabla}T_\mathrm{e}\right).
\end{equation}
As opposed to the first method, we compute the Biermann current on the cell edges, using for the gradients the two end vertices, where $n_{\rm e}$ and $T_{\rm e}$ are calculated beforehand by averaging the cell-centred values of the 8 surrounding cells (Fig.~\ref{fig:improved}). The corresponding EMF component in the $x$-direction is thus defined as
\begin{equation}
E_{\mathrm{B},x}^{i,j+\sfrac{1}{2},k+\sfrac{1}{2}} = \frac{c k_\mathrm{B}}{e} \left( 
\overline{T_\mathrm{e}} \frac{\left[ \ln n_\mathrm{e} \right]}{\Delta x} +  
\frac{\left[ T_\mathrm{e} \right]}{\Delta x} 
\right),
\end{equation}
where the edge temperature is defined as the average of the two end vertex temperatures as
\begin{equation}
\overline{T_\mathrm{e}} = 
\frac{1}{2} \left( T_\mathrm{e}^{i+\sfrac{1}{2},j+\sfrac{1}{2},k+\sfrac{1}{2}} + T_\mathrm{e}^{i-\sfrac{1}{2},j+\sfrac{1}{2},k+\sfrac{1}{2}} \right),
\end{equation}
and the two undivided differences are also computed using the two end vertex log densities and temperatures as
\begin{equation}
\left[ \ln n_\mathrm{e} \right] = \ln n_\mathrm{e}^{i+\sfrac{1}{2},j+\sfrac{1}{2},k+\sfrac{1}{2}}- \ln n_\mathrm{e}^{i-\sfrac{1}{2},j+\sfrac{1}{2},k+\sfrac{1}{2}} ,
\end{equation}
and
\begin{equation}
\left[ T_\mathrm{e} \right] = 
T_\mathrm{e}^{i+\sfrac{1}{2},j+\sfrac{1}{2},k+\sfrac{1}{2}}- T_\mathrm{e}^{i-\sfrac{1}{2},j+\sfrac{1}{2},k+\sfrac{1}{2}}.
\end{equation}
This approach has the benefit of ensuring a zero field at machine precision in the isothermal case. In this case, the Biermann electric field is indeed proportional to the gradient of a scalar function, so that its curl exactly vanishes. Thanks to the Stokes theorem, we can exactly preserve this property at the discrete level. This results into what is usually called a `structure-preserving scheme'. This alternative method, called the `improved' method in this paper, is motivated by the fact that the photo-ionized IGM tends to be globally isothermal in cosmological simulations. It should be noted that our method slightly differs from the method proposed by \cite{Graziani2015}, who decided to preserve the Rankine-Hugoniot relations at shocks instead. Note also that the refinement of the edge-defined EMFs is automatically and suitably carried out by the code \citep{Teyssier2006,Fromang2006}.

%%%%%%%%%%%%%%%%%%%%%%%%%%%%%%%%%%%%%%%%%%%%%%%%%%

\section{Idealised tests}
\label{sect:tests}

In this section, we first present the numerical tests that we carry out to validate the newly-implemented Biermann battery module. Then, two complete suites of astrophysical tests are detailed, namely the propagation of a Strömgren sphere and a Sedov blast wave. They act as a sandbox where the most important mechanisms in the ultimate cosmological simulations of the EoR are separately investigated, that is to say propagating ionization fronts and shock waves.

\subsection{Static smooth tests}

We perform here simple tests of the Biermann battery solver alone, in a static configuration and for a single time step. This allows us to validate our new scheme with known analytical solutions. We focus on the differences between the naive and the improved method. 

\subsubsection{Biermann battery with orthogonal gradients}

\begin{figure}
\centering
\includegraphics[width=8.5cm]{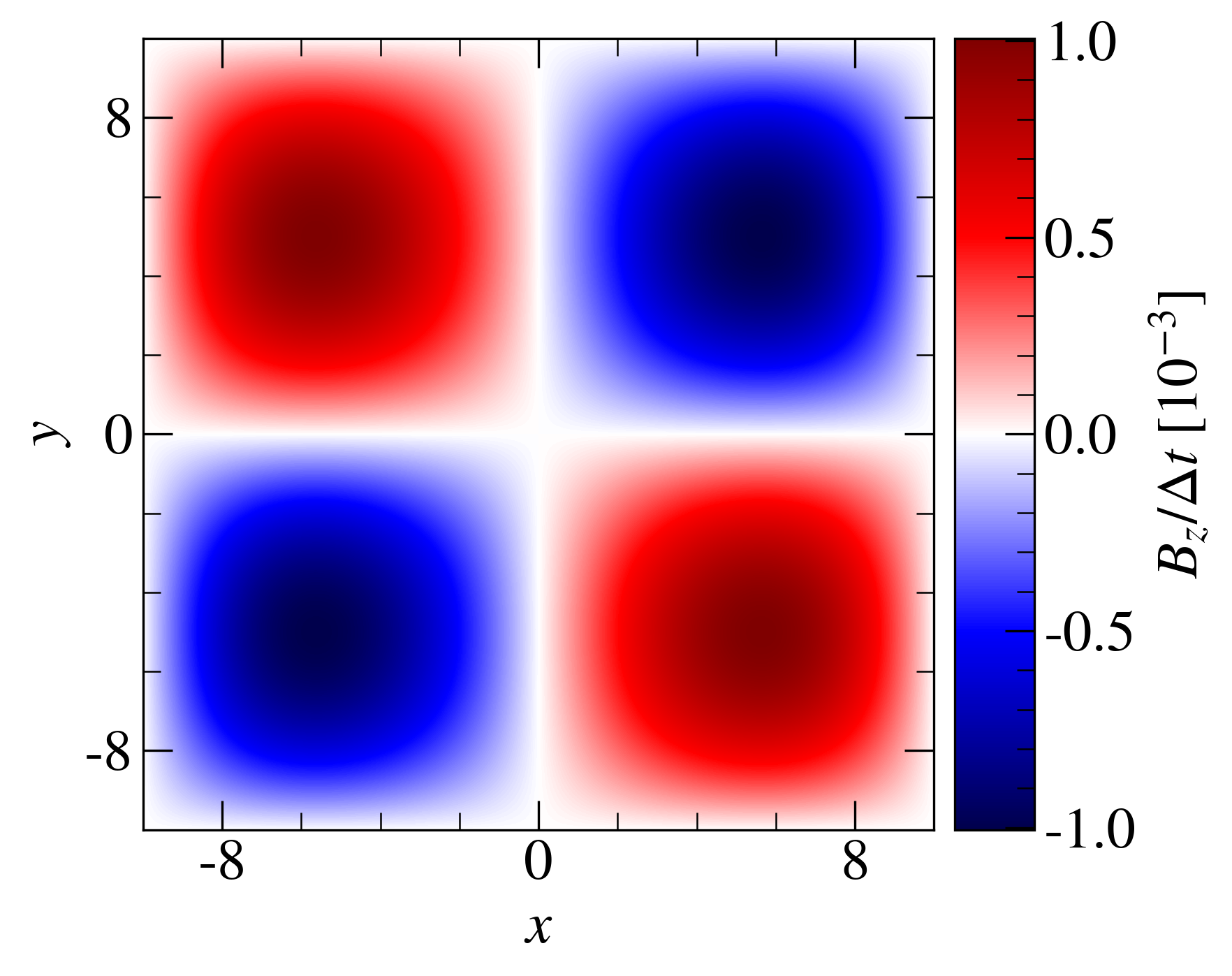}
\caption{Map of the temporal derivative of the magnetic field after one computational step in the smooth flow test. Both methods yield the same result.}
\label{fig:smoothB}
\end{figure}

\begin{figure}
\centering
\includegraphics[width=8.5cm]{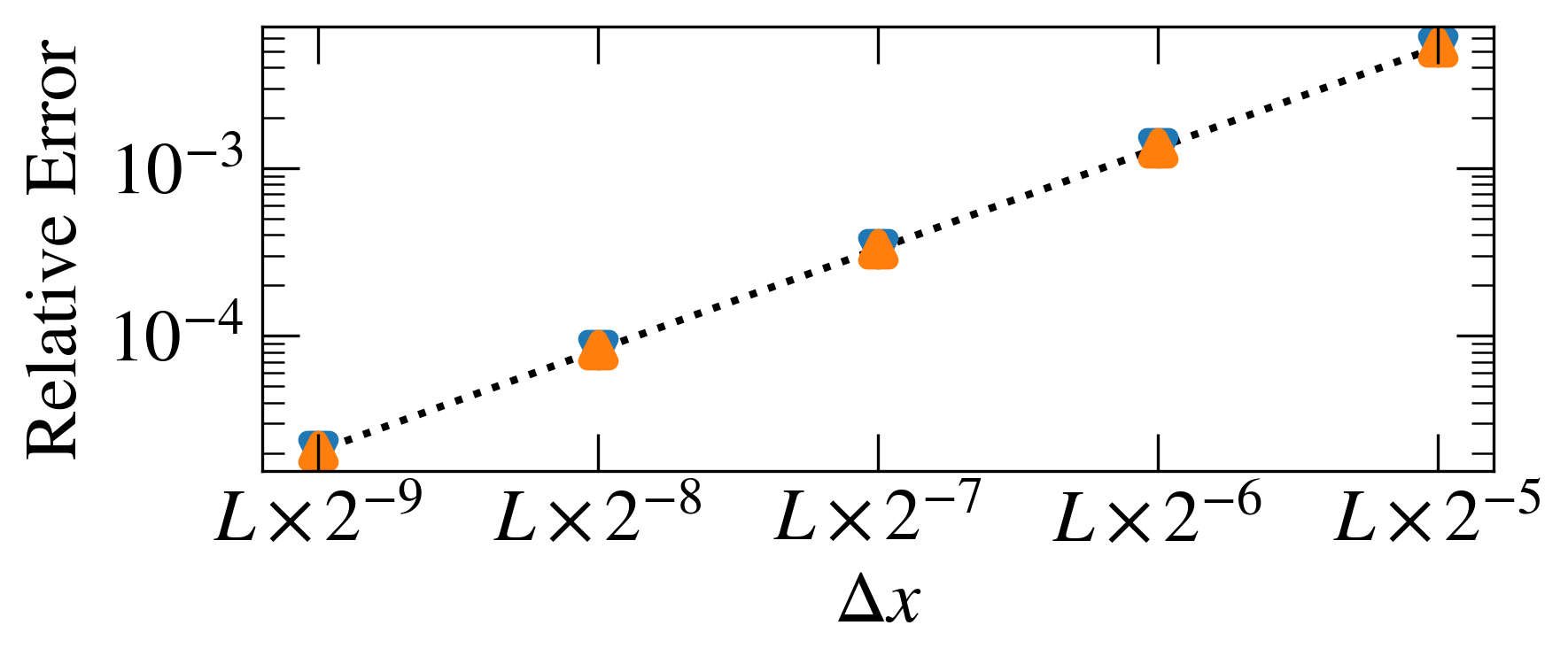}
\caption{$\mathcal{L}^1$ norm of the relative error to the analytical solution of the smooth flow test as a function of decreasing resolution ($L$ represents the box size). Both methods (blue: naive, orange: improved) yield the same result and follow the dotted line of slope 2.}
\label{fig:smoothError}
\end{figure}

This first test problem proposed originally by \cite{Toth2012} and \cite{Graziani2016} considers smooth electron density and pressure profiles with orthogonal gradients so that a strong Biermann-induced magnetic field is created with a known analytical solution. For simplicity, we use dimensionless units where $c=e=k_{\rm B}=1$. The test assumes an initial condition of $\boldsymbol{B}=\boldsymbol{0}$, $\boldsymbol{u}=\boldsymbol{0}$, $n_\mathrm{e}=n_0+n_1\cos\left(k_xx\right)$, and $p_\mathrm{e}=p_0+p_1\cos\left(k_yy\right)$, where $k_x=k_y=\pi/10$, $n_0=p_0=1$, and $n_1=p_1=0.1$. Consequently, $n_{\rm e}$ and $p_{\rm e}$ gradients are indeed orthogonal, which maximises the effect of the Biermann battery. The computational box spans a square domain from $-10$ to $10$ in both $x$- and $y$-directions. The resulting magnetic field in the $z$-direction after one time step is given by the analytical form
\begin{equation}
\frac{B_z}{\Delta t}=-\frac{k_xk_yn_1p_1\sin\left(k_xx\right)\sin\left(k_yy\right)}{\left(n_0+n_1\cos\left(k_xx\right)\right)^2}.
\label{eq:smooth}
\end{equation}
We use a resolution of $128 \times 128$ points and a single time step to recover the analytic solution. Our two methods (naive and improved) yield the same result (Fig.~\ref{fig:smoothB}), and both agree with Eq.~(\ref{eq:smooth}) within a relative error of $\sim 3 \times 10^{-4}$ for this resolution.

Exploiting the smooth character of this problem, we are able to perform a convergence study by varying the resolution of the grid and by computing the $\mathcal{L}^1$ norm of the error compared to the analytical solution. Figure~\ref{fig:smoothError} confirms that the results of the two methods are quite similar in this case, and also shows that the error decreases with the mesh size as $\Delta x^2$, typical of standard second-order schemes.

\subsubsection{Biermann battery under isothermal conditions}

\begin{figure}
\centering
\includegraphics[width=8.5cm]{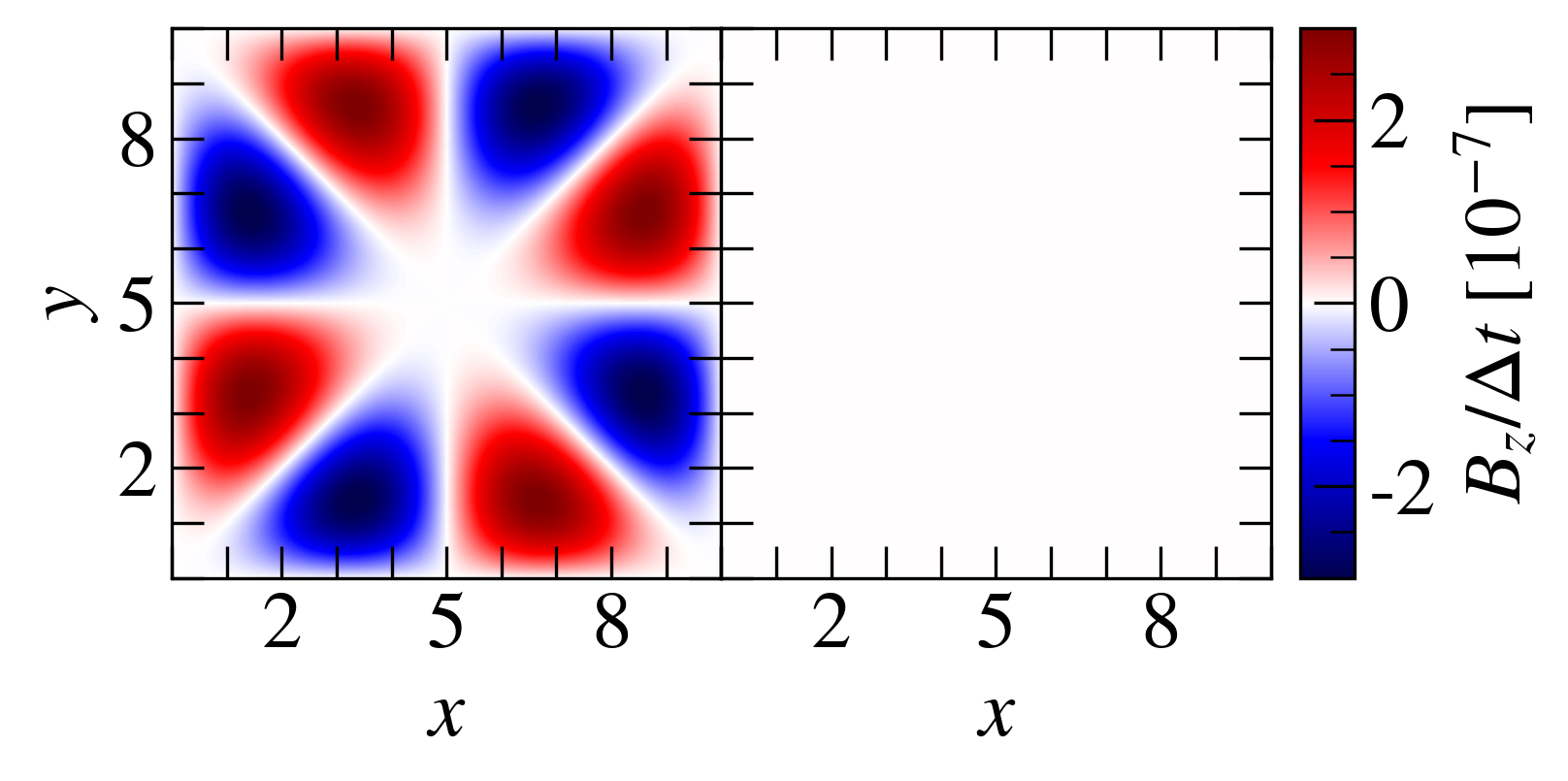}
\caption{Map of the temporal derivative of the magnetic field after one computational step in the isothermal flow test for the naive (left) and improved (right) methods. The improved map only features machine-round-off errors around 0, which are not visible on this scale.}
\label{fig:isoth}
\end{figure}

The second test is similar to the previous one with the difference that the initial density and pressure profiles are now parallel to each other with
\begin{equation}
n_\mathrm{e}=n_0+n_1\left(\sin\left(k_xx\right)^2+\sin\left(k_yy\right)^2\right),
\end{equation}
\begin{equation}
p_\mathrm{e}=p_0+p_1\left(\sin\left(k_xx\right)^2+\sin\left(k_yy\right)^2\right).
\end{equation}
This corresponds to strict isothermal conditions. Because both density and pressure gradients are now parallel, we do not expect any Biermann-battery-generated magnetic field. The resulting maps for $B_z/\Delta t$ after one computational step are shown in Fig.~\ref{fig:isoth}. Note that we still use dimensionless units here. The naive method generates a spurious magnetic field at the level of the truncation errors in the calculation of the curl of the pressure gradient, around $10^{-7}$ in these units. The improved method, on the other hand, generates vanishingly small values at the level of double precision numerical round-off errors, around $10^{-16}$ in these units. This proves the benefit of the improved method in preventing the formation of a spurious, residual magnetic field in isothermal conditions where no battery is expected.

\subsection{Strömgren sphere test}
\label{sect:stromgren}

\begin{figure}
\centering
\includegraphics[width=8.5cm]{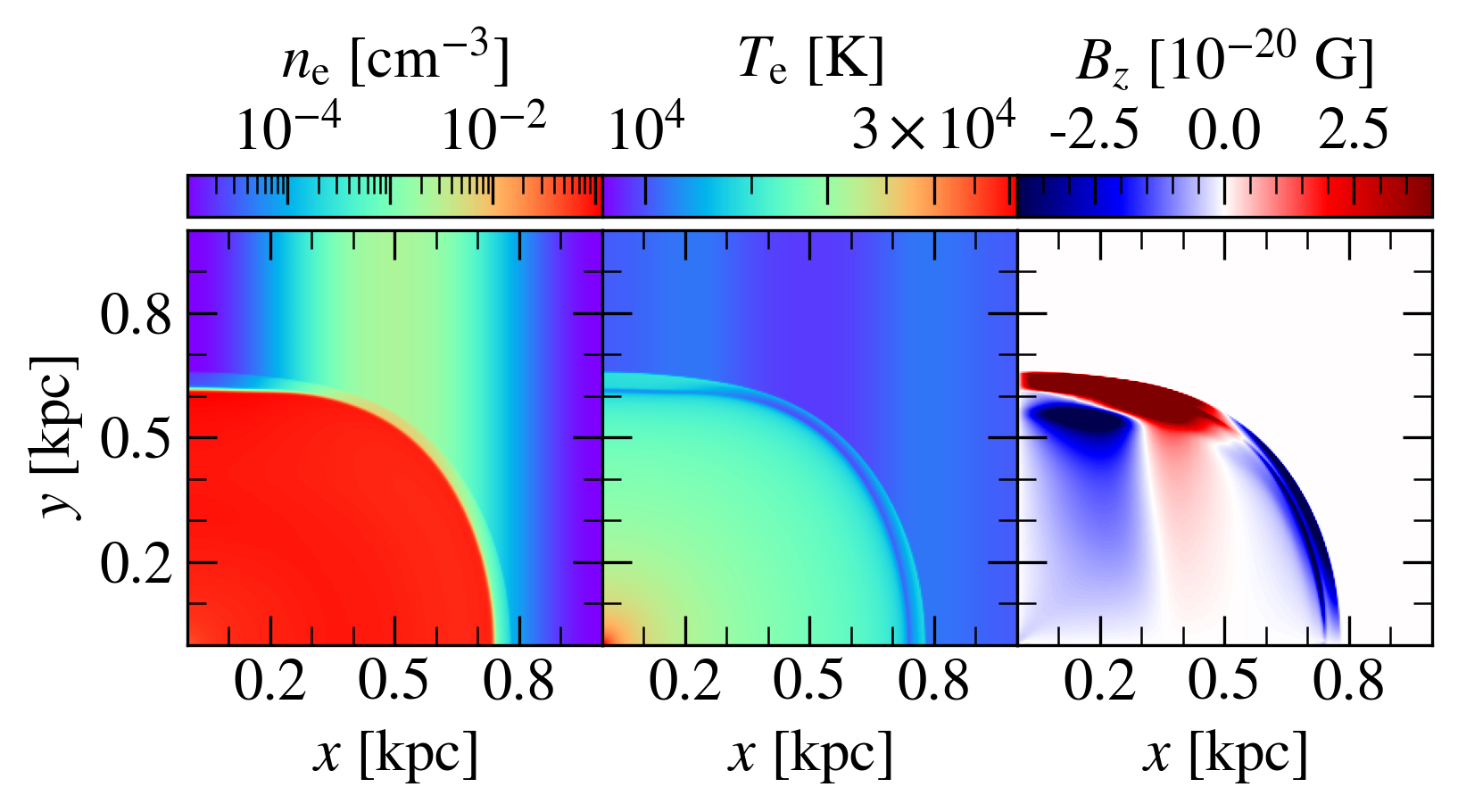}
\caption{Maps of the electron density (left panel), electron temperature (middle panel) and vertical (out of plane) magnetic field (right panel) at $t=15$~Myr for the \textsc{str-20\%-i} simulation.}
\label{fig:str_map}
\end{figure}

\begin{figure*}
\centering
\includegraphics[width=17cm]{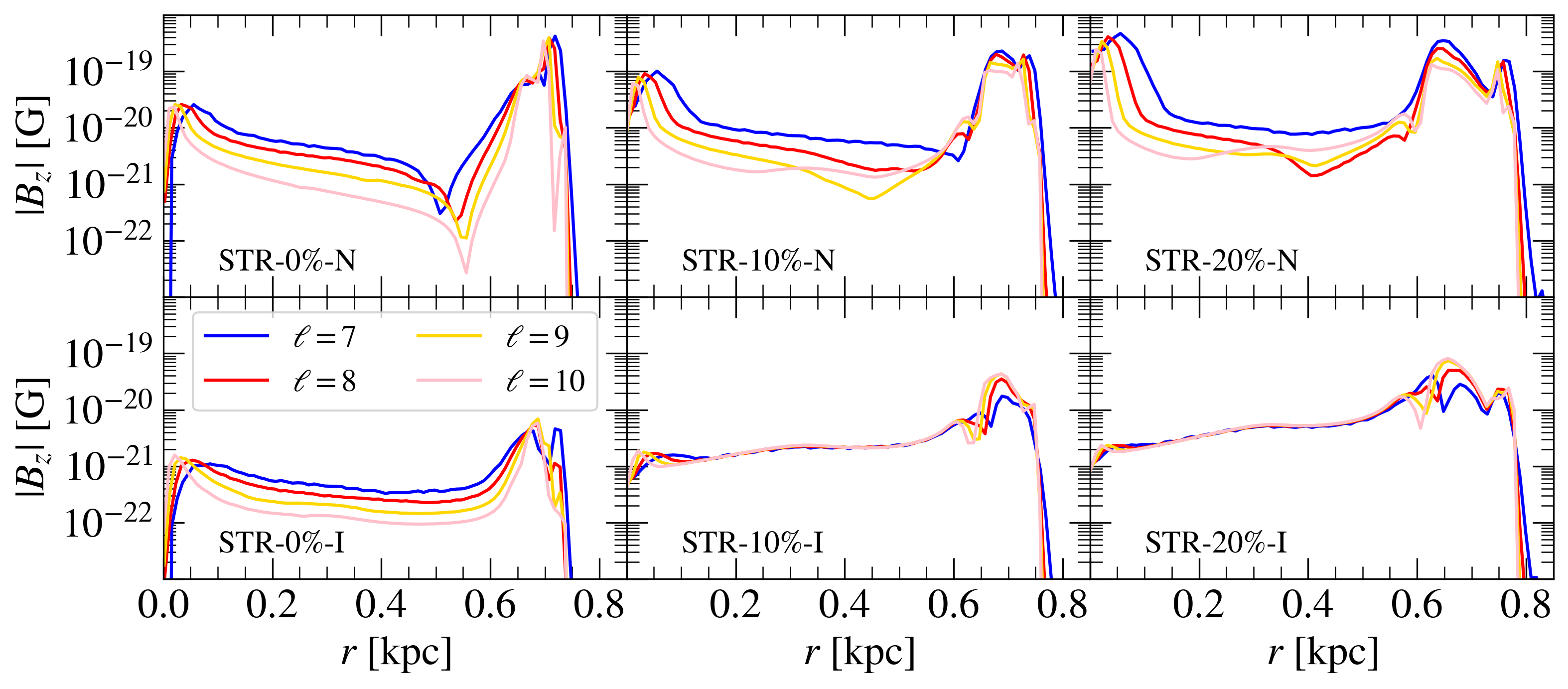}
\caption{Vertical magnetic field profiles at the end of the Strömgren sphere test simulations. Each colour represents a different resolution, as labelled in the bottom left panel.}
\label{fig:str_prof}
\end{figure*}

We now investigate the effect of the Biermann battery in a situation that is relevant for cosmological simulations, namely the propagation of a Strömgren sphere. This test and the one presented in the following section are particularly challenging for any Biermann battery implementation, because they feature strong discontinuities that could result, as explained in \cite{Graziani2015}, in the so-called Biermann catastrophe. The propagation of a Strömgren sphere is also the basic process governing the Biermann battery within ionization fronts during the EoR \citep[e.g.][]{Subramanian1994,Gnedin2000}. Note that from now on we switch to physical units.

The Strömgren sphere simulations are performed in a two-dimensional (2D) square box of size $L \times L$ without AMR refinements, where $L=1$ kpc. The mesh resolution is $\Delta x=L\times2^{-\ell}$, where $\ell$ is the refinement level. A source emitting ionizing photons of energy 29.6 eV at a rate of $\dot{N}=2 \times 10^{28}~\rm{s}^{-1}~{\rm cm}^{-1}$ is placed at the origin, namely the lower left corner of the box. In order to create favourable conditions for the Biermann battery, the initial gas density profile is spatially modulated using $n=n_0+n_1\cos\left(k_xx\right)$, with $k_x=2\pi/L$, while the initial temperature profile is kept uniform with $T=T_0$. The gas is composed of pure hydrogen of cross-section $1.6 \times 10^{-18}$ cm$^2$.

Our initial conditions are similar to the physical conditions in the ISM in a typical galaxy, namely $n_0=0.1$~cm$^{-3}$ and $T_0=10^4$~K. We perform a series of simulations in which we vary the amplitude of the density modulation and the numerical method used (Table~\ref{tab:str_sim}). The simulation time is fixed to 15~Myr.

\begin{table}
\centering
\begin{tabular}{ c c c }
\hline
 Simulation name & $n_1$ & Method \\ 
\hline
 \textsc{str-0\%-n} & 0.0 $\times n_0$ & Naive \\  
 \textsc{str-10\%-n} & 0.1 $\times n_0$ & Naive \\  
 \textsc{str-20\%-n} & 0.2 $\times n_0$ & Naive \\  
 \textsc{str-0\%-i} & 0.0 $\times n_0$ & Improved \\  
 \textsc{str-10\%-i} & 0.1 $\times n_0$ & Improved \\  
 \textsc{str-20\%-i} & 0.2 $\times n_0$ & Improved \\  
\hline
\end{tabular}
\caption{Table of the simulation parameters used in the Strömgren sphere test runs.}
\label{tab:str_sim}
\end{table}

Results of this test for the simulation with 20\% modulation and the improved method (called here \textsc{str-20\%-i}) are shown in Fig.~\ref{fig:str_map}. The Strömgren sphere is clearly visible as the fully-ionized red bubble in the electron density map (left panel of Fig.~\ref{fig:str_map}). The initial density modulation can also be seen in the same panel as a blue-green sine wave in the background. This modulation causes a clear distortion of the sphere in the $x$-direction, transforming it into an ellipsoid.
Two discontinuities are present in this flow: the ionization front itself and a dynamical shock launched into the surrounding ISM. This rather complex structure is more visible in the temperature map (middle panel of Fig.~\ref{fig:str_map}). Looking at the range of the colour map, though, we can convince ourselves that the temperature evolution is quite close to being isothermal. 

The propagation of the Strömgren sphere and its accompanying shock generate a magnetic field of a few $10^{-20}$~G in the vertical, $z$-direction. We can also see a clear modulation of the magnetic field inside the Strömgren sphere, whose sign reflects the initial modulation of the density field. The magnitude of this generated field can be estimated using a simple dimensional analysis of Eq.~(\ref{eq:battery}), with
\begin{equation}
\label{eq:Bamp}
B_\mathrm{S}=\frac{c}{e}\frac{k_\mathrm{B} T_\mathrm{S}}{r_\mathrm{S}^2} t_\mathrm{S},
\end{equation}
where $T_{\rm S}$, $r_{\rm S}$ and $t_{\rm S}$ are characteristic temperature, length and time scales. Using $t_{\rm S}=15$ Myr as the simulation time, $r_{\rm S}=0.7$ kpc as the approximate radius of the Strömgren sphere and $T_{\rm S}=10^4$~K, we get $B_{\rm S} \simeq 10^{-20}$ G, in very good agreement with our results.

In Fig.~\ref{fig:str_prof}, we show the radial profiles of the vertical magnetic field generated in our 6 different Strömgren sphere simulations. The upper panels are for the naive method, while the lower panels show the results of the improved method. The left column corresponds to the case without any density modulation, for which density and pressure gradients are supposed to be both strictly radial with no Biermann battery. Unfortunately, both methods produce a spurious residual field in this case because of numerical truncation errors, in particular due to the mapping of a spherical profile onto a Cartesian grid. Note however that the improved method produces a spurious field that is more than 50 $\times$ smaller than the naive method, which is a significant improvement. In both cases, the spurious field decreases with increasing resolution, except close to the source at the origin and at the ionization front. This is a problem similar to the one reported by \cite{Graziani2015}, where such singularities and discontinuities in the flow are particularly difficult to handle by the Biermann battery term. Note that the improved method performs significantly better than the naive method close to these singularities. 

In the middle and right columns of Fig.~\ref{fig:str_prof}, we show the results obtained for the 10\% and 20\% modulation. We can clearly see that the improved method is now converging towards a unique profile, with more difficulties at the ionization front. The naive method, on the other hand, is far from converging. It is clear that, in this case, the spurious field strength is always comparable to the correct field strength, making the naive method totally unfit for this test. Only at the highest resolution $\ell=10$ and for the largest modulation does the naive method show signs of convergence towards what seems to be the correct solution.

\subsection{Sedov blast wave test}
\label{sect:sedov}

\begin{figure}
\centering
\includegraphics[width=8.5cm]{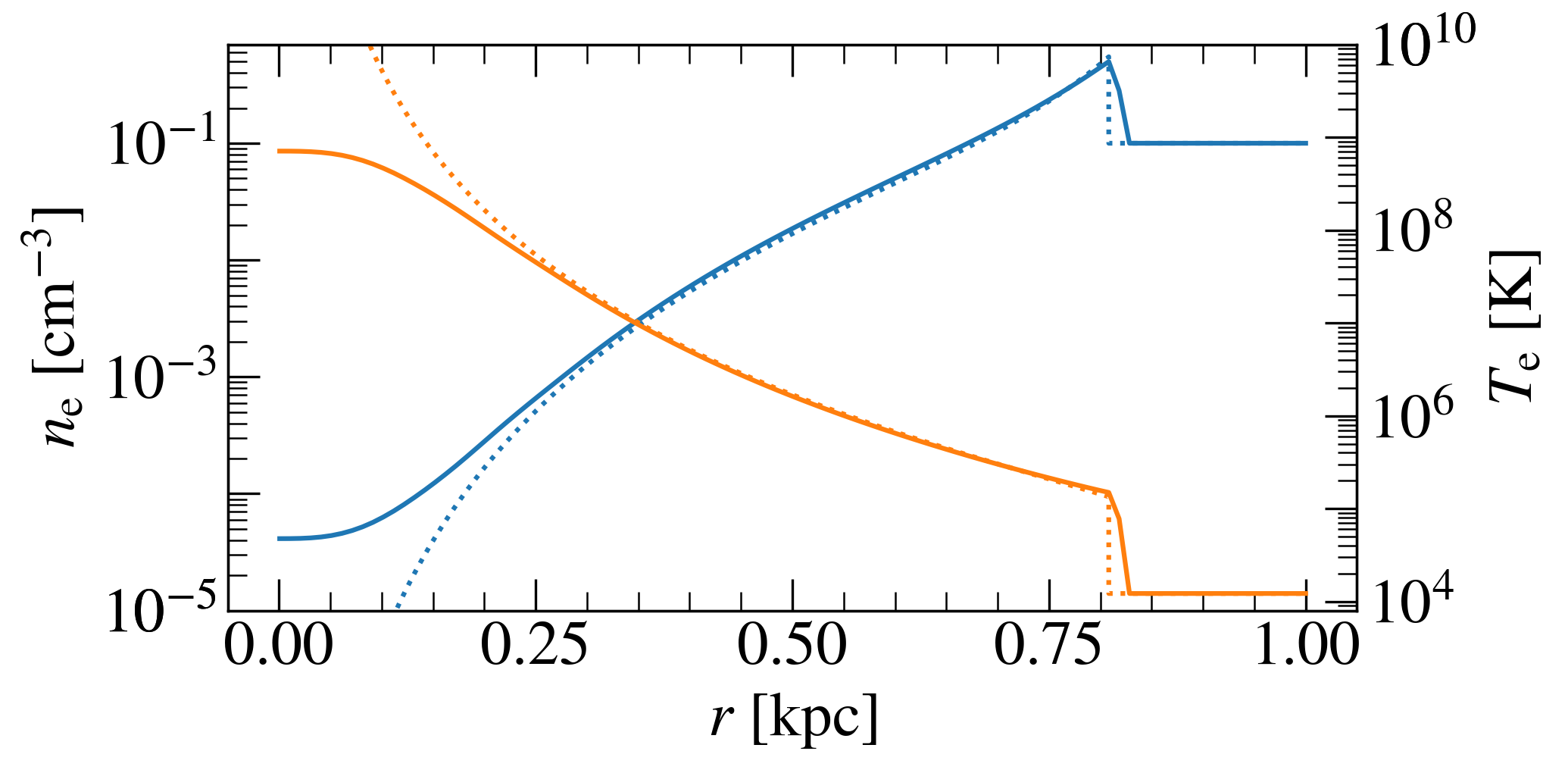}
\caption{Electron density (blue) and temperature (orange) profiles at $t=5$~Myr for the Sedov blast wave test with no modulation (the \textsc{sed-0\%-i} simulation) and spatial resolution $\ell=10$. The exact analytical Sedov solutions are over-plotted as dotted curves.}
\label{fig:sed_jump}
\end{figure}

\begin{figure}
\centering
\includegraphics[width=8.5cm]{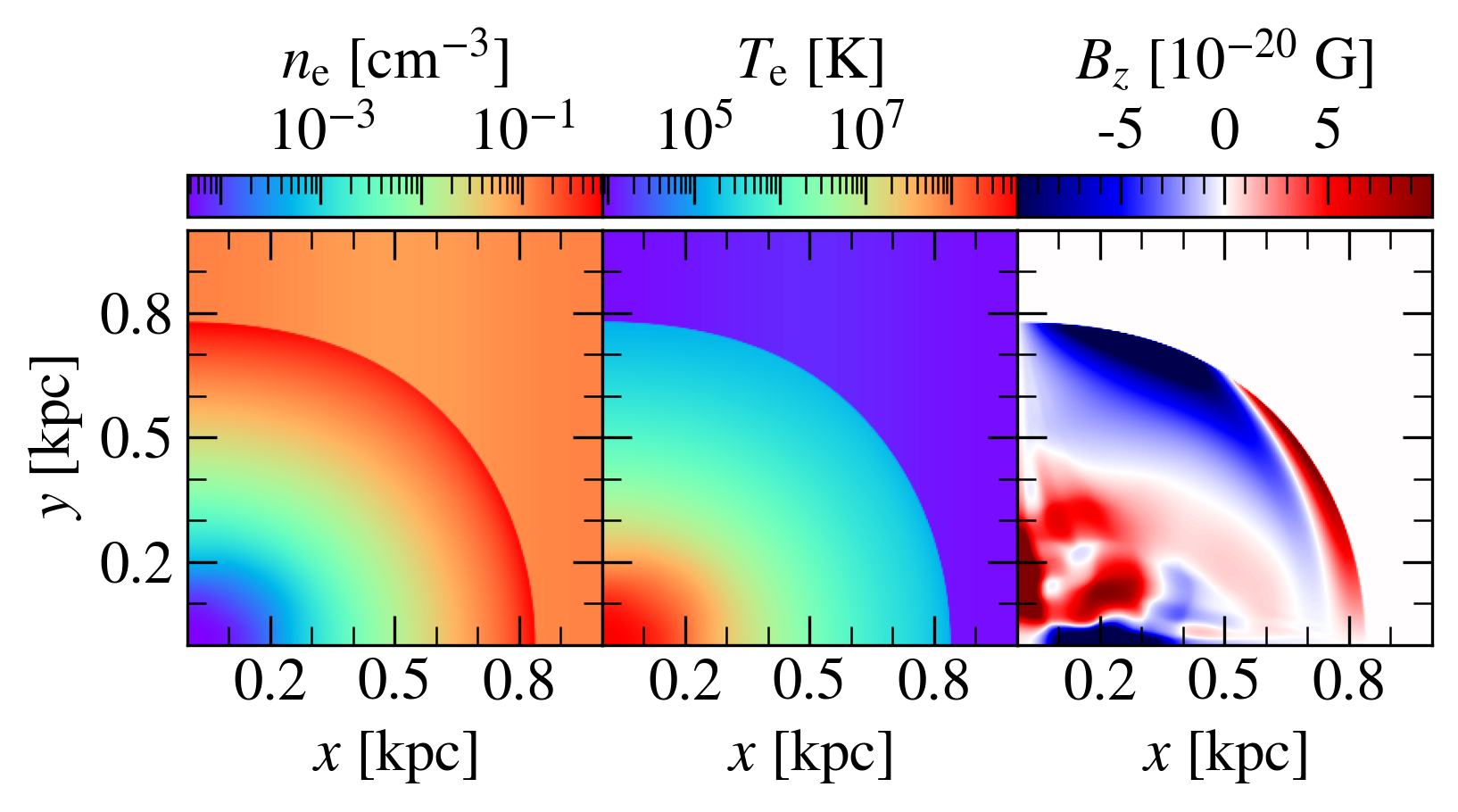}
\caption{Maps of the electron density (left panel), electron temperature (middle panel) and vertical magnetic field (right panel) at $t=5$~Myr for the Sedov blast wave test with 20\% modulation (the \textsc{sed-20\%-i} simulation) and spatial resolution $\ell=10$.}
\label{fig:sed_map}
\end{figure}

\begin{figure*}
\centering
\includegraphics[width=17cm]{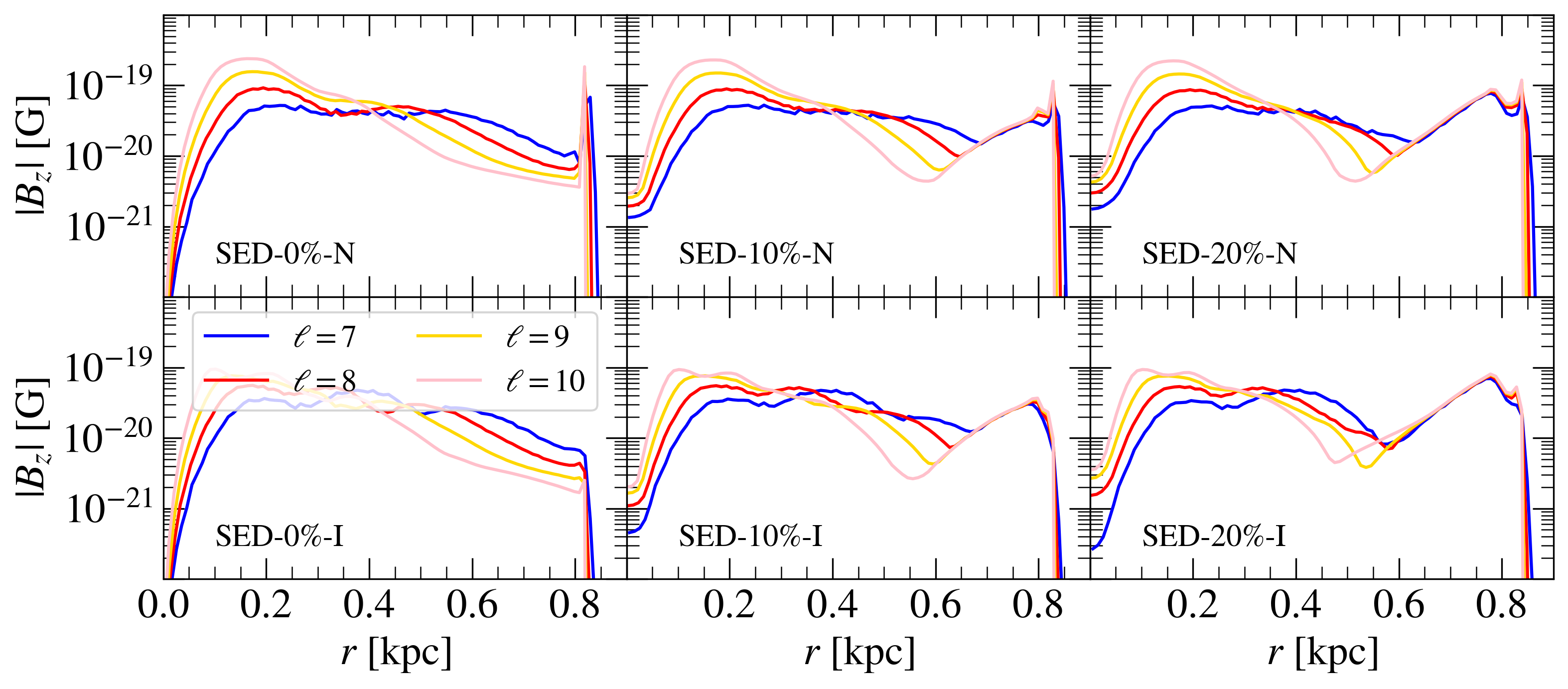}
\caption{Vertical magnetic field profiles at $t=5$~Myr for our 6 different Sedov blast wave test simulations. Each colour represents a different resolution, as labelled in the bottom left panel.}
\label{fig:sed_prof}
\end{figure*}

In this section, we perform a second series of tests relevant for the Biermann battery in shock waves \citep[e.g.][]{Kulsrud1997,Davies2000}. We consider the well-known case of a 2D Sedov blast wave. While the Strömgren sphere maintains a quasi-isothermal regime, the Sedov blast wave exhibits large temperature variations with sharp discontinuities, which cause numerical difficulties, especially across shock fronts \citep{Fatenejad2013,Graziani2015}.

We proceed in a very similar manner to the previous section. Our computational domain is a 2D square box with a width of 1~kpc. Here again we do not use any AMR refinements. A modulated initial density profile is used in order to generate a non-spherical shock with a proper Biermann battery: $n=n_0+n_1\cos\left(k_xx\right)$ and $T=T_0$ ($n_0$ and $T_0$ have the same values as in the previous section). An initial energy pulse of $3.2 \times 10^{50}$ erg/pc is placed at the origin so as to have a blast wave similar to a cluster of 1000 combined SN explosions. The adiabatic index of the gas is set to $\gamma=1.4$. We use the same terminology to identify our different simulations as for the Strömgren sphere tests (i.e. as shown in Table \ref{tab:str_sim}). Contrary to the previous section, the gas is assumed to be always fully ionized. The simulation time is set to 5~Myr.

We first check that our numerical solution without modulation and with resolution $\ell=10$ reproduces the analytical Sedov solution (Fig.~\ref{fig:sed_jump}). The density jump at the shock front is close to 6, as predicted by the Rankine-Hugoniot relations, and the density and temperature profiles nicely follow the analytical solution. Note however that a significant deviation is observed in the centre. This is due to the initial singularity only captured by one cell at $t=0$. After the blast wave has expanded to its final radius at $t=5$~Myr, this single cell is affecting the entire central region within a radius as large as $r=0.35$~kpc.

When we apply the initial 10\% or 20\% density modulation, the spherical symmetry of the propagating shock wave is broken into an elliptical shock (Fig.~\ref{fig:sed_map}). A vertical magnetic field of strength up to $10^{-19}$ G is generated in the post-shock region. We can motivate such a value using again a dimensional analysis with Eq.~(\ref{eq:Bamp}), yielding a predicted value of $B_{\rm S} \simeq 2 \times 10^{-19}$ G, using $t_{\rm S}=5$ Myr, $r_{\rm S}=0.8$ kpc, and $T_{\rm S}=10^6$ K.

Figure~\ref{fig:sed_prof} compares our 6 different simulations. The left column shows the case without modulation, for which no magnetic field should be generated as the profiles are purely radial. Here again, the improved method performs better than the naive one, although not as spectacularly as for the Strömgren sphere test. A noticeable difference is however visible right after the shock front. The naive method clearly suffers from the phenomenon described in \cite{Graziani2015} as the Biermann catastrophe, for which the spurious magnetic field generated within the shock discontinuity increases when one increases the resolution. The improved method, on the contrary, shows a much better behaviour, with the spurious field decreasing with increasing resolution, similar to the other improved method proposed by \cite{Graziani2015}.

The middle and right panels of Fig.~\ref{fig:sed_prof} show the profile of the vertical magnetic field for the modulated cases. The two methods seem to converge towards what we believe is the correct solution, with the improved method doing a slightly better job than the naive one. The central region within $0.35$~kpc is however not converging at all in both cases. We believe this is due to the initial singularity generating stronger and stronger spurious fields in the centre, as the temperature becomes higher, ultimately diverging in the very centre. As the resolution is increased, the problematic central region shrinks, but this occurs at a relatively slow rate. This also leads to a spurious magnetic field morphology in the centre of the explosion (Fig.~\ref{fig:sed_map}).

To conclude this section, we would like to stress that these tests are particularly difficult as they feature many singularities and discontinuities, as well as a very modest deviation from spherical symmetry. In a more realistic three-dimensional environment, we expect shock waves and ionization fronts to exhibit stronger non-parallel density and pressure gradients and a less ambiguous Biermann battery process. Note however that our improved method produces significantly better results that the naive method, especially across shock fronts or in nearly isothermal conditions. It allows us to see converged results for the two test suites we have proposed, although this convergence appears to be surprisingly difficult to achieve.

%%%%%%%%%%%%%%%%%%%%%%%%%%%%%%%%%%%%%%%%%%%%%%%%%%

\section{Cosmological simulations}
\label{sect:cosmo}

In this section, we present the cosmological simulations we run to study the Biermann battery at play during the EoR. First, we describe the global framework in which these simulations are carried out, namely the \sphinx{} project, including the employed initial conditions. Then, we focus on the results of the multiple cosmological simulations.

\subsection{The SPHINX framework}
\label{sect:sphinx}

For our cosmological simulations we use the same setup and sub-grid models as in the \sphinx{} suite of simulations presented in \cite{Rosdahl2018}. We refer to that paper for details, but briefly recap the main ingredients here.

We run cosmological volumes of size 2.5 and 5.0 co-moving Mpc (cMpc) starting at $z=150$ and ending at $z=6$. The $\Lambda$CDM initial conditions are generated with \textsc{Music} \citep{Hahn2011} using cosmological parameters consistent with the Planck 2013 results \citep{Ade2014}. We use hydrogen and helium mass fractions of $X=0.76$ and $Y=0.24$, respectively, and the gas has an initial metal mass fraction of $Z_{\rm init}=6.4\times 10^{-6}$ in order to have high enough cooling rates to form the first stars at $z \simeq 15$ in the absence of molecular hydrogen formation, which is not modelled.

The number of DM particles in the 2.5 and 5 cMpc volumes are $128^3$ and $256^3$ respectively, giving a mass resolution of $\mdm = 2.6 \times 10^5 \ \msun$ in both volumes. We consider a DM halo to be resolved if it contains 300 particles or more, meaning we resolve halo masses $\gtrsim 8 \times 10^7 \ \msun$. The largest halo masses at $z=6$ in the smaller and larger volumes are $\sim 5 \times 10^9 \ \msun$ and $\sim 10^{10} \ \msun$ respectively\footnote{The halo mass, $M_{\rm vir}$, is defined as the total DM mass within an ellipsoidal virialised region and identified with the \adaptahop{} halo finder \citep{Tweed2009}}. The gas cells (which are also used for the ionizing radiation field and gravitational potential from all matter) are adaptively refined, splitting a cell in 8 if it contains 8 DM particles or an equivalent mass in gas and stars, or if the local Jeans length is smaller than 4 cell widths. The maximum number of such hierarchical splits from the coarse resolution is 8, so the coarse cell width is 19.6 ckpc (2.8 kpc at $z=6$) and the minimum cell width reached in the ISM of galaxies is 76.3 cpc (10.9 pc at $z=6$). 

Stellar particles, each representing a population of stars with a mass of $10^3 \ \msun$, are formed out of the gas if 1) the hydrogen density $\nh>10 \ \cci$, 2) the turbulent Jeans length is smaller than one cell width, and 3) the gas is locally convergent. The gas is converted into stellar particles with an efficiency that varies with the local virial parameter and turbulence \citep{Kimm2017,Trebitsch2017,MartinAlvarez2020a}. The star formation is typically very bursty, as the local star formation efficiency can reach values even higher than unity under the right conditions.

Each stellar particle produces a series of individual $10^{51}$ erg type II SN explosions that are stochastically sampled between $3$ and $50$ Myr of its lifetime. An average mass fraction $\eta_{\rm SN}=0.2$ of the particle is ejected back into the surrounding gas (the exact fraction depending on how many SN explosions the particle produces), with a metal yield of $y=0.075$. We assume an average SN progenitor mass of $m_{\rm SN}=5 \ \msun$, meaning each particle produces an average of 40 SN explosions, which is about 4 times more than for a \citet{Kroupa2001} IMF. This boost in the number of SN explosions is necessary to suppress galaxy growth and reproduce the UV luminosity function of galaxies in larger \sphinx{} volumes \citep[see detailed discussion in][]{Rosdahl2018}. SN explosions are implemented with a mechanical feedback model that injects the predicted final momentum of the snowplow phase if the Sedov--Taylor phase is not resolved, but a pre-snowplow kick otherwise \citep{2014ApJ...788..121K}.

Ionizing radiation is injected into the grid cells from stellar particles, using age- and metallicity-dependent luminosities derived from the Binary Population and Spectral Synthesis code (\bpass) version 2.2.1 \citep[][]{Eldridge2007, Stanway2018}, assuming an initial mass function with slopes of $-1.3$ from $0.1$ to $0.5 \ \msun{}$ and $-2.35$ from $0.5$ to $100 \ \msun$. We use the M1 moment method to advect the radiation, using the variable speed of light approximation \citep{Katz2017} to reduce the computational cost. The speed of light at the highest AMR resolution, corresponding to the ISM of galaxies, is $0.0125 \ c$, but the radiation speeds up to $20 \%$ of the real speed of light at the lowest AMR resolution, corresponding to cosmological voids. The Biermann battery being sensitive to the electron number density, we are not affected by post-reionization artefacts in the residual neutral fraction due to the reduced speed of light framework \citep{Ocvirk2019}. Moreover, \citet{Deparis2019} showed that adopting a value of 10\% of the speed of light delayed complete reionization by no more than $\Delta z=0.5$ in redshift, with respect to a full speed of light simulation, making it compatible with current observational knowledge of the timing of large-scale reionization \citep[e.g.][]{Kulkarni2019}. Three groups of radiation are tracked: a low-energy group bracketed by the ionization energies for H (13.6 eV) and He \textsc{i} (24.6 eV), an intermediate group bracketed by the ionization energies for He \textsc{i} and He \textsc{ii} (54.4 eV), and a high-energy group only bracketed on the lower side by the He \textsc{ii} ionization energy.

Hydrogen and helium thermochemistry, coupled with the local radiation field via photo-ionization, and heating are evolved with the method described in \cite{Rosdahl2013}. The ionization fractions of hydrogen and helium are evolved in each gas cell and advected with the gas. For $T>10^4$ K, metal-line cooling is computed with \cloudy{}-generated tables \citep[][version 6.02]{Ferland1998}. Fine structure cooling rates from \cite{Rosen1995} are used in the $T \le 10^4$ K regime, allowing the gas to radiatively cool to a density-independent temperature floor of $15$ K.

The simulations also solve in parallel the ideal MHD equations using the Godunov scheme and a CT method \citep{Teyssier2006,Fromang2006}. Only the induction equation (Eq.~\ref{eq:induc}) is modified as we add the Biermann source term to the EMF. The system of equations is formulated using `supercomoving variables' to account for the expansion of the Universe in the FLRW metric \citep{Martel1998}. In doing so, only one additional source term, involving the scale factor $a$, is introduced in the induction equation, which conveniently leaves all the other MHD equations unchanged \citep[more details are provided in][]{Rieder2017}. The initial magnetic field is identically zero everywhere. Note that these computations are done in double precision.

\subsection{Results}
\label{sect:results}

\begin{table}
\centering
\begin{tabular}{ c c c c c c }
\hline
 Simulation name & $L$ & $\ell_{\rm min}$ & $\ell_{\rm max}$ & Method & Comment \\ 
  & (cMpc) & & & & \\
\hline
 \textsc{spx2.5-n} & 2.5 & 7 & 14 & Naive & \\  
 \textsc{spx2.5-i} & 2.5 & 7 & 14 & Improved & \\  
 \textsc{spx2.5-i-nosn} & 2.5 & 7 & 14 & Improved & SNe off \\  
 \textsc{spx5.0} & 5.0 & 8 & 15 & Improved &  \\  
\hline
\end{tabular}
\caption{Parameters of our various cosmological simulations. $L$ is the box size. $\ell_{\rm min}$ and $\ell_{\rm max}$ respectively refer to the minimum and to the maximum refinement level.}
\label{tab:sph_sim}
\end{table}

\begin{figure*}
\centering
\includegraphics[width=17cm]{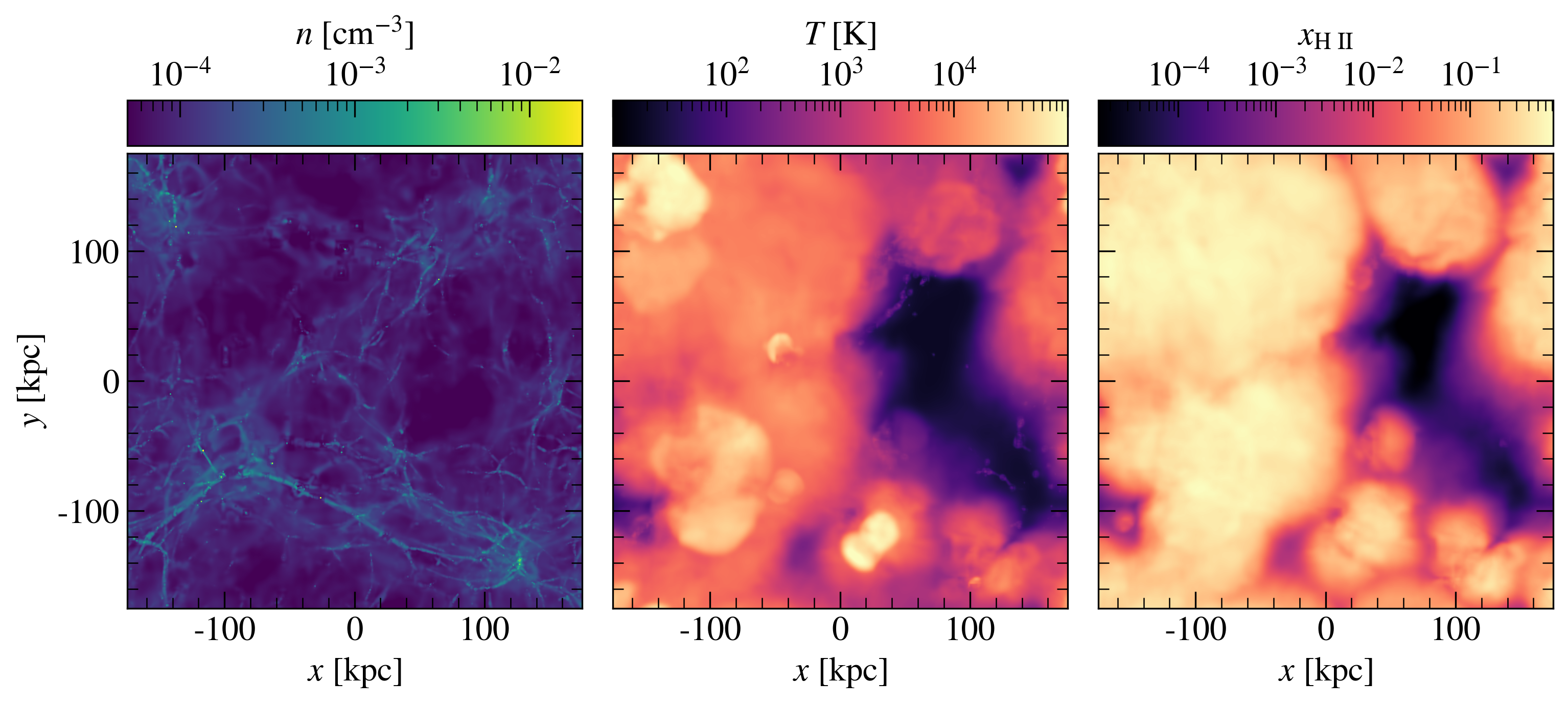}
\caption{Mass-weighted maps projected along the $z$-direction for the density (left), temperature (centre), and ionization fraction (right) at the end ($z=6$) of the \textsc{spx2.5-i} simulation.}
\label{fig:sph25_dTx}
\end{figure*}

\begin{figure}
\centering
\includegraphics[width=8.5cm]{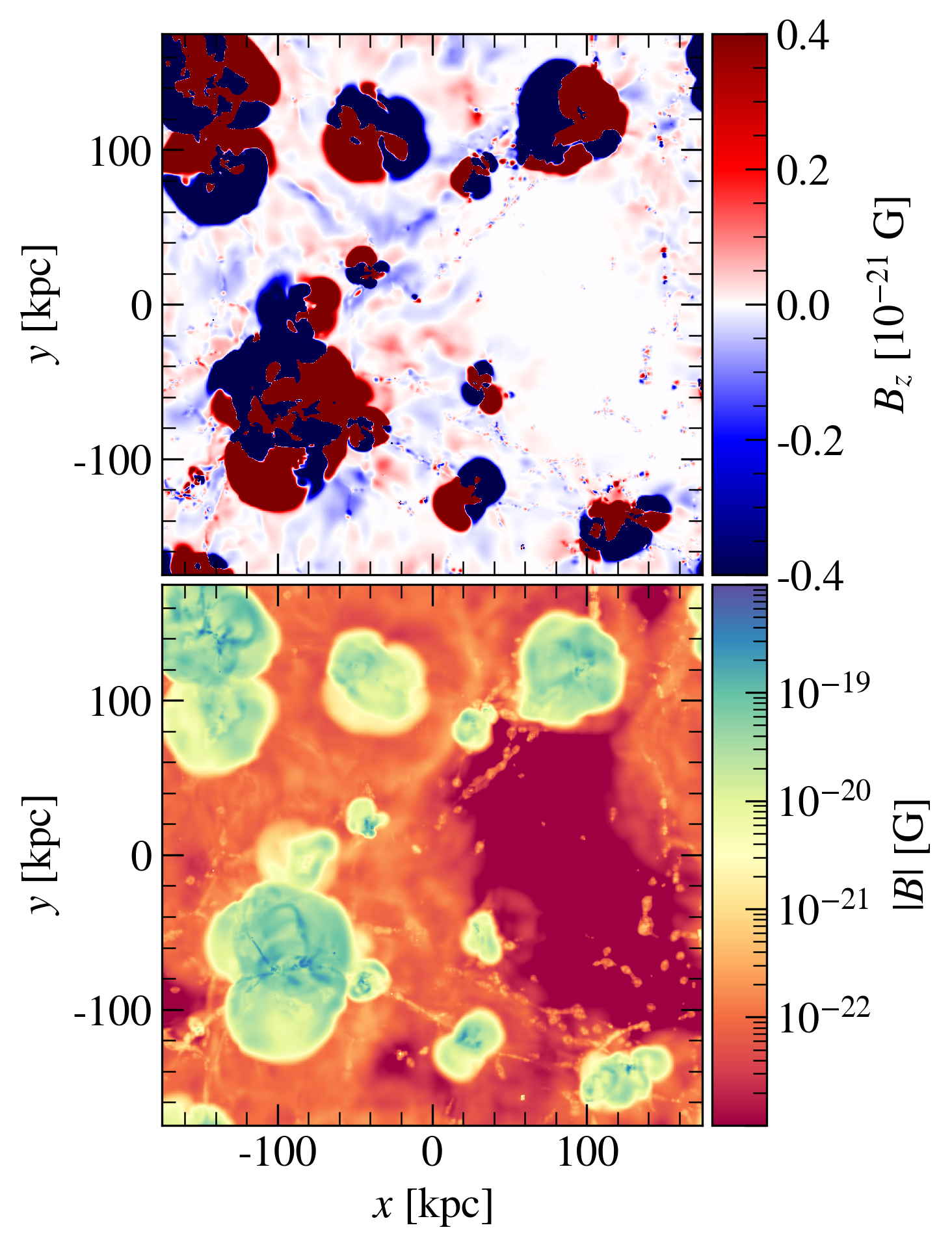}
\caption{Mass-weighted maps projected along the $z$-direction for $B_z$ (top) and magnetic field strength (bottom) at the end ($z=6$) of the \textsc{spx2.5-i} simulation.}
\label{fig:sph25_B}
\end{figure}

\begin{figure}
\centering
\includegraphics[width=8.5cm]{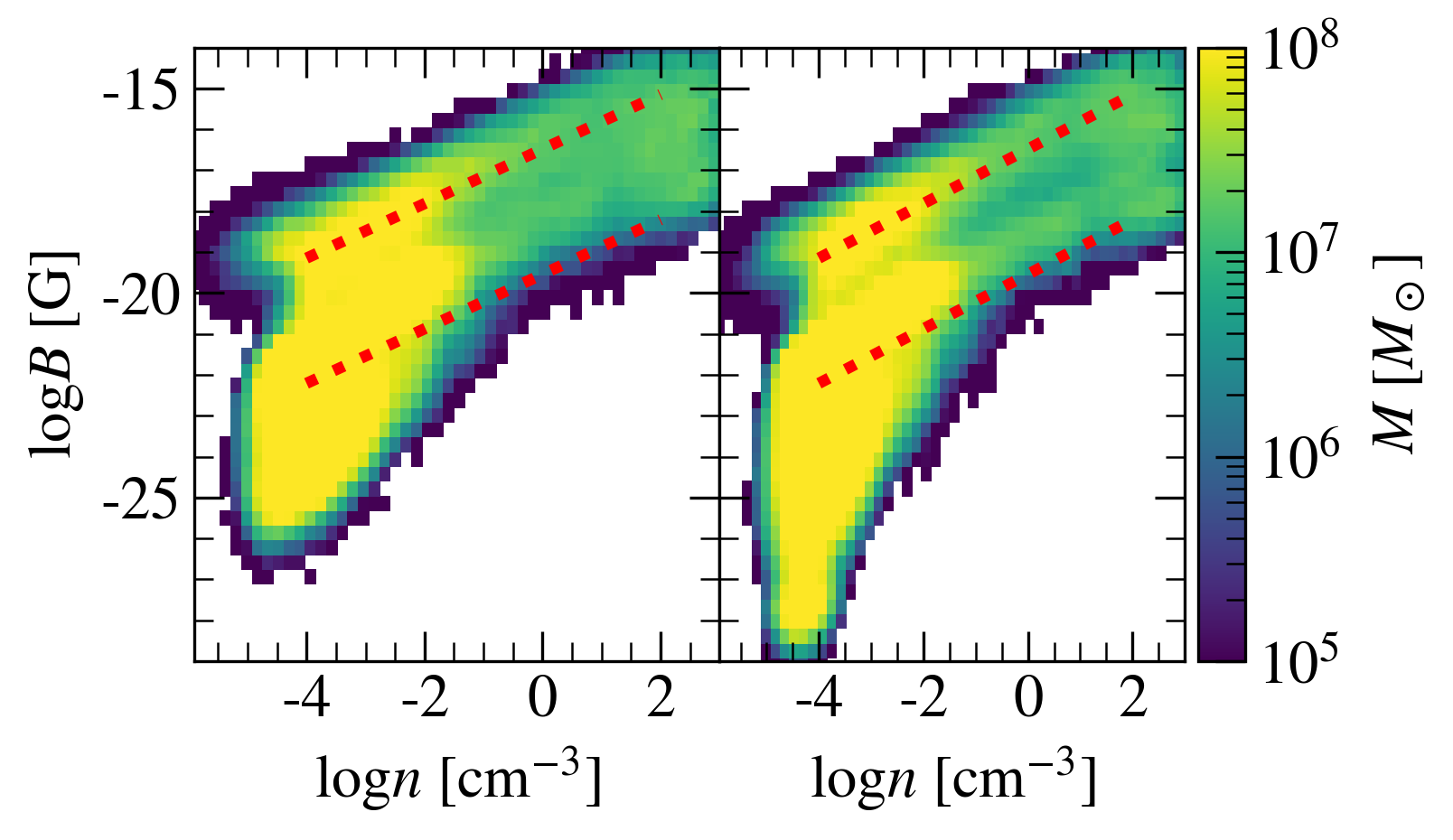}
\caption{Mass-weighted 2D histograms of the magnetic field versus density at the end ($z=6$) of the \textsc{spx2.5-n} (left) and \textsc{spx2.5-i} (right) simulations. The two analytical $n^{2/3}$ relations are present in red dotted lines.}
\label{fig:sph25_Bd}
\end{figure}

In this section, we present the results of our cosmological simulations using radiation-magneto-hydrodynamics and our new implementation of the Biermann battery. Table \ref{tab:sph_sim} summarises the parameters used for the various runs. We explore the effect of both the naive, less accurate, Biermann battery scheme and the improved one. We also check the influence of the box size on our results. We also perform one simulation with SN feedback turned off.

\subsubsection{Three modes of magnetic field generation}

We report here on the results of our fiducial \textsc{spx2.5-i} run, featuring the improved Biermann battery scheme, SN feedback, and a volume of 2.5 cMpc on the side. Figure~\ref{fig:sph25_dTx} presents the final snapshot of the simulation at $z=6$. The gas distribution shows a typical filamentary structure surrounding large voids. The reionization of the IGM is almost complete by that time. The whole box is nearly fully ionized except for a small lower-density patch in which ionization fronts have not had time yet to percolate. The temperature of the gas in ionized regions is about $10^4$~K. Note however the high temperature bubbles ($10^5 - 10^6$~K) in the vicinity of the largest haloes due to feedback from SN-driven explosions.

The generated magnetic field is presented in Fig.~\ref{fig:sph25_B}. The upper panel shows the $z$-component of the magnetic field. All three components follow a similar pattern, namely a complicated morphology alternating between positive and negative values imprinted by the complex shape of the density and temperature gradients. The magnetic field is generally oriented towards the cross-product of these gradients, broadly encircling H \textsc{ii} regions \citep{Gnedin2000}. 

In the lower panel, the magnetic field strength map reveals the co-existence of three different regimes: 1) a very low magnetic field strength with $B \simeq 10^{-24}$~G in voids filled with neutral hydrogen, 2) a low, intermediate regime with $B \simeq 10^{-20}$~G in ionized regions, and finally 3) higher values $B \simeq 10^{-18}$~G in a regime associated with SN-driven hot bubbles. 

The first mode is associated with the Biermann battery process during the linear evolution of gas density perturbations, subject to the complex thermal history after recombination \citep{Naoz2013}.

The second regime is associated with ionized regions and the classical Biermann battery channel, in which the IGM is progressively magnetised as ionization fronts spread. The process can be depicted as propagating Strömgren spheres gradually filling the whole volume with $10^{-20}$~G fields, as illustrated in Sect. \ref{sect:stromgren} with the Strömgren sphere test. 

The third mode can be seen as large bubbles hosting magnetic fields two orders of magnitude higher than the previous case. These bubbles are localised in the vicinity of star-forming regions. In fact, they precisely lie within the above-mentioned high-temperature bubbles, suggesting that this third regime is concentrated around massive haloes where SN explosions occur. At this redshift, these SN-driven bubbles are far from having percolated. This third magnetic field generation channel is not associated with ionization fronts but with the propagation of galactic winds, in which magnetic fields are generated within SN-driven shocks, in a similar way to our Sedov blast wave test in Sect. \ref{sect:sedov}. It owes its higher amplitude ($10^{-18}$~G) to the higher temperature of the gas. We will discuss this point later below.

We have also compared the results of our two Biermann battery implementations, the naive and the improved one, in the density-magnetic field phase-space diagram shown in Fig.~\ref{fig:sph25_Bd}. We can clearly see here again the three modes we have just introduced. The lower branch, where the magnetic field strength is lower than $10^{-23}$~G, corresponds to the linear regime prior to the EoR \citep{Naoz2013}. We will discuss more the origin of these very weak fields in Sect.~\ref{sect:volume}. 

The upper branch shows the classical behaviour of ideal MHD with the field strength scaling as $n^{2/3}$ (shown as the dashed lines). Noticeably, we observe the presence of two distinct and parallel sub-branches that both scale as $n^{2/3}$. This bi-modal distribution is due to the co-existence of the ionization generation channel and of the shock generation channel. 

When comparing the naive method (left panel of Fig.~\ref{fig:sph25_Bd}) to the improved method (right panel of Fig.~\ref{fig:sph25_Bd}), we see a very good qualitative agreement, supporting the robustness of our conclusions. Quantitatively, however, the field strength in the lower branch (associated with the linear regime of the Biermann battery) is overestimated by the naive method, owing to its largest truncation errors, as revealed in the tests section.

To support this interpretation, we would like to analytically derive the amplitude of the Biermann battery generated fields using here again Eq.~(\ref{eq:Bamp}). For the ionization front branch, the length scale $r_{\rm S}$ can be estimated as the radius of the Strömgren sphere emanating from a typical simulated galaxy with
\begin{equation}
r_\mathrm{S}=\left(\frac{3\dot{N}}{4\pi\beta_\mathrm{rec}\bar{n}^2}\right)^{1/3},
\end{equation}
where $\beta_{\rm rec} \simeq 2.6 \times 10^{-13}$ cm$^3$ s$^{-1}$ is the recombination rate at $T=10^4$~K and $\bar{n}$ is the mean baryon density at $z=6$, namely $\bar{n} \simeq 10^{-4}$ cm$^{-3}$. We consider for the ionizing photon rate $\dot{N}=10^{50}$ s$^{-1}$, corresponding to the luminosity of our largest simulated halo, after accounting for the proper escape fraction \citep{Rosdahl2018}. The resulting Strömgren sphere size is $r_{\rm S} \simeq 70$ kpc, consistently with Fig.~\ref{fig:sph25_B}. For $t_{\rm S}$, we use the age of the Universe at the end of the simulation with $t_{\rm S}=1$ Gyr and for $T_{\rm S}$, we use the ionized IGM temperature with $T_{\rm S}=10^4$ K. Eq.~(\ref{eq:Bamp}) hence yields an amplitude of $B_{\rm low} \simeq 10^{-22}$ G for the mean density of the Universe $\bar{n}$.

\begin{figure}
\centering
\includegraphics[width=8.5cm]{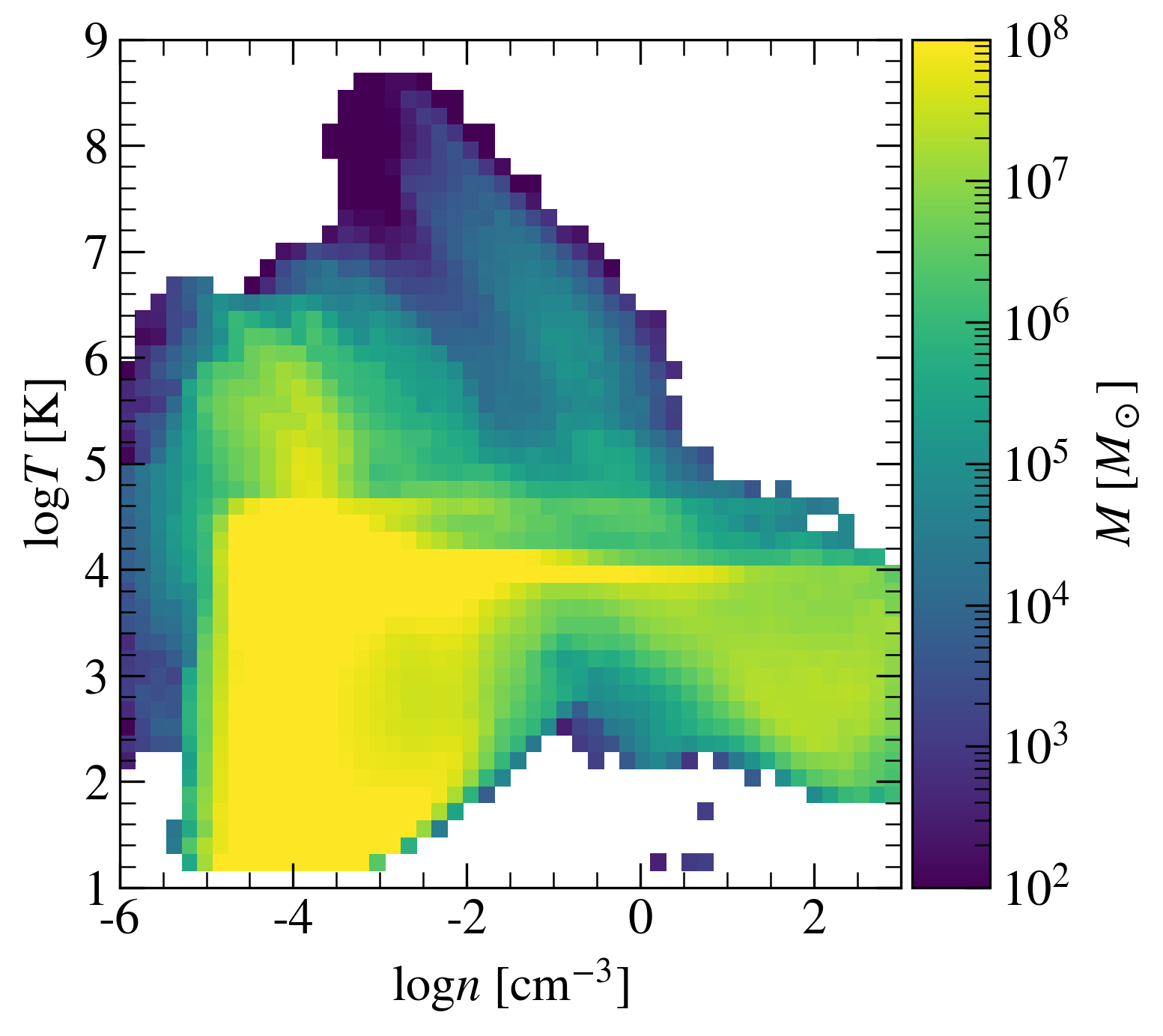}
\caption{Mass-weighted 2D histogram of the temperature versus density at the end ($z=6$) of the \textsc{spx2.5-i} simulation.}
\label{fig:sph25_Td}
\end{figure}

As for the high-$B$ branch, the spatial extent of the hot magnetised bubbles is dictated by galactic winds that spread the fields generated in SN-driven shocks. To fully describe analytically such a process would be beyond the scope of the current paper \citep[see][for example]{Bertone2006}. We consider that a good approximation is to use the observed characteristic radius of the hot bubbles $r_{\rm S}=20$ kpc (see Fig.~\ref{fig:sph25_B}). The most substantial change between the two regimes is the electron temperature. An examination of the temperature distribution (Fig.~\ref{fig:sph25_Td}) shows that the typical temperature in the shocked regions is $T_{\rm S}=10^6$ K at the mean density in the Universe, which, using here again Eq.~(\ref{eq:Bamp}), translates into an amplitude of $B_{\rm high} \simeq 10^{-19}$~G at $\bar{n} \simeq 10^{-4}$~cm$^{-3}$. 

The previous two analytical estimates are then used to normalise the two branches as $B=B_{\rm low} \left(n/\bar{n}\right)^{2/3}$ and $B=B_{\rm high} \left(n/\bar{n}\right)^{2/3}$. We show the resulting predictions as the two red dashed lines in Fig.~\ref{fig:sph25_Bd}. We see that the $n^{2/3}$ scaling is not strictly reproduced by our simulation results. This is usually attributed to turbulent processes providing additional magnetic field amplification, on top of bare gravitational contraction \citep{Dubois2008,Quilis2020}. This relatively weak effect can be interpreted as the very first sign of a galactic dynamo (see Discussion below).

\begin{figure}
\centering
\includegraphics[width=8.5cm]{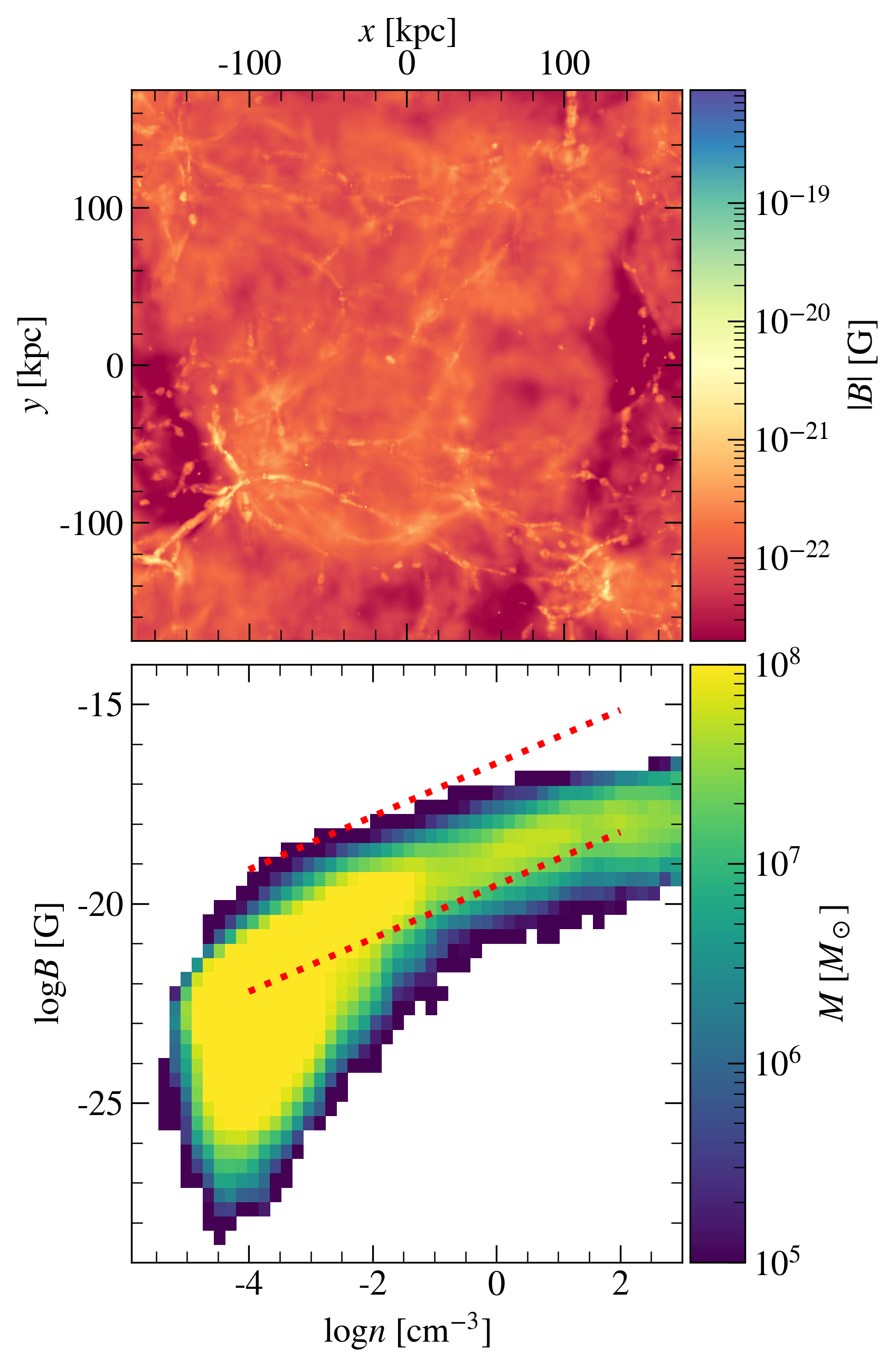}
\caption{Map of the mass-weighted magnetic field strength projected along the $z$-direction (top) and 2D mass histogram of the magnetic field versus density (bottom) at $z=6$ for the \textsc{spx2.5-i-nosn} simulation. The two analytical relations derived in the paper are shown as the red dashed lines.}
\label{fig:sph25_nosn}
\end{figure}

In order to demonstrate that the high-$B$ branch is indeed generated by SN-driven galactic winds, we run the same simulation but disabling SN feedback (\textsc{spx2.5-i-nosn}). The hot magnetised bubbles that are previously seen around massive haloes completely disappear (Fig.~\ref{fig:sph25_nosn}). We confirm this by inspecting the phase-space diagram, from which the high-$B$ branch also disappears, only leaving the ionization branch and the linear regime. This confirms that this new magnetic field generation channel directly originates from SN-driven galactic winds. Note that in the early simulations of \cite{Gnedin2000}, galactic winds were not as strong as they are now in our newest cosmological models, since they did not include explosions of massive stars. As a consequence, this magnetic field generation channel was missing.

\subsubsection{Finite volume effects}
\label{sect:volume}

\begin{figure}
\centering
\includegraphics[width=8.5cm]{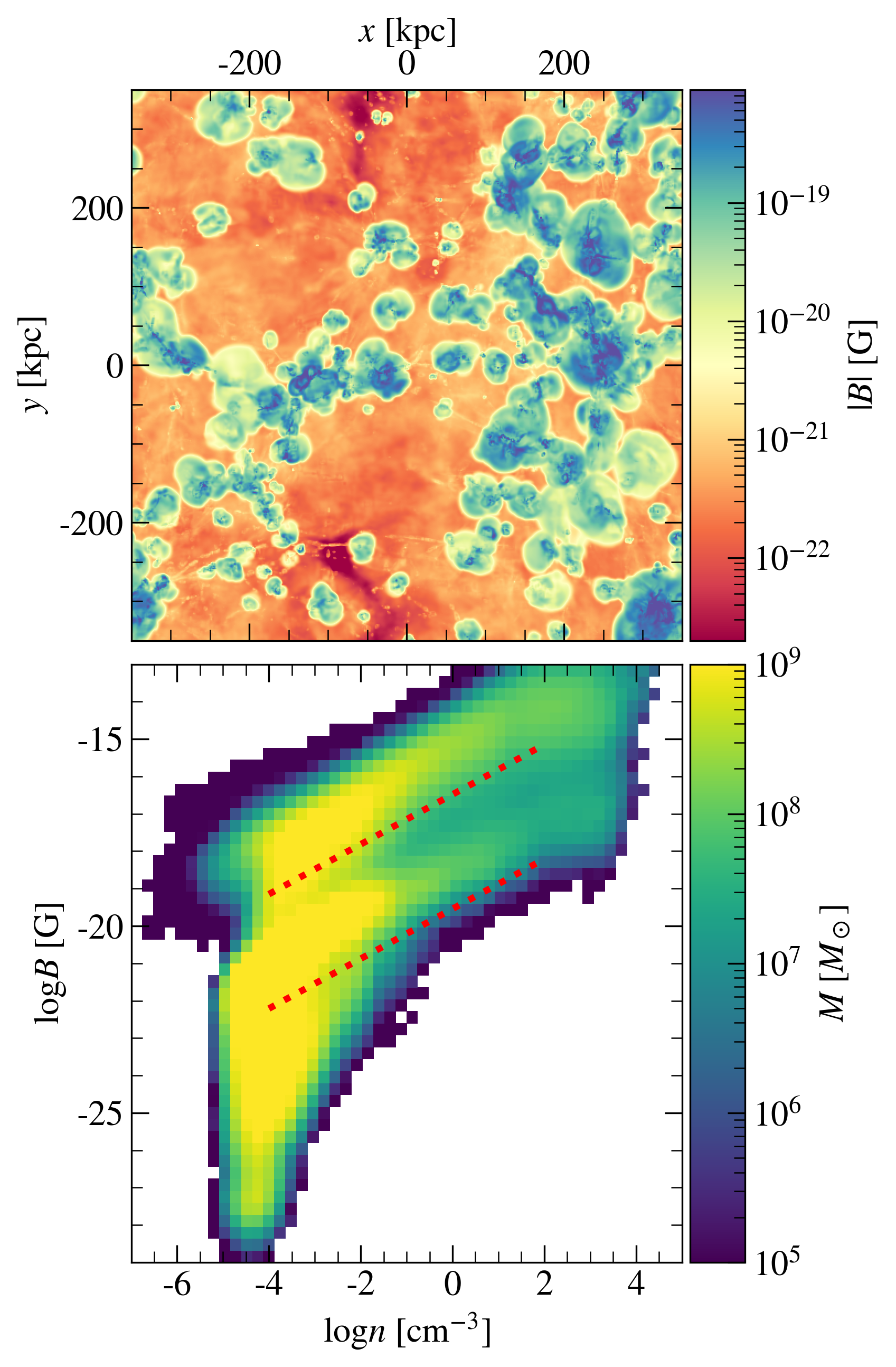}
\caption{Map of the mass-weighted magnetic field strength projected along the $z$-direction (top) and 2D mass histogram of the magnetic field versus density (bottom) at $z=6$ for the large box \textsc{spx5.0} simulation. The two analytical relations derived in the text are shown as the red dashed lines.}
\label{fig:sph50}
\end{figure}

\begin{figure}
\centering
\includegraphics[width=8.5cm]{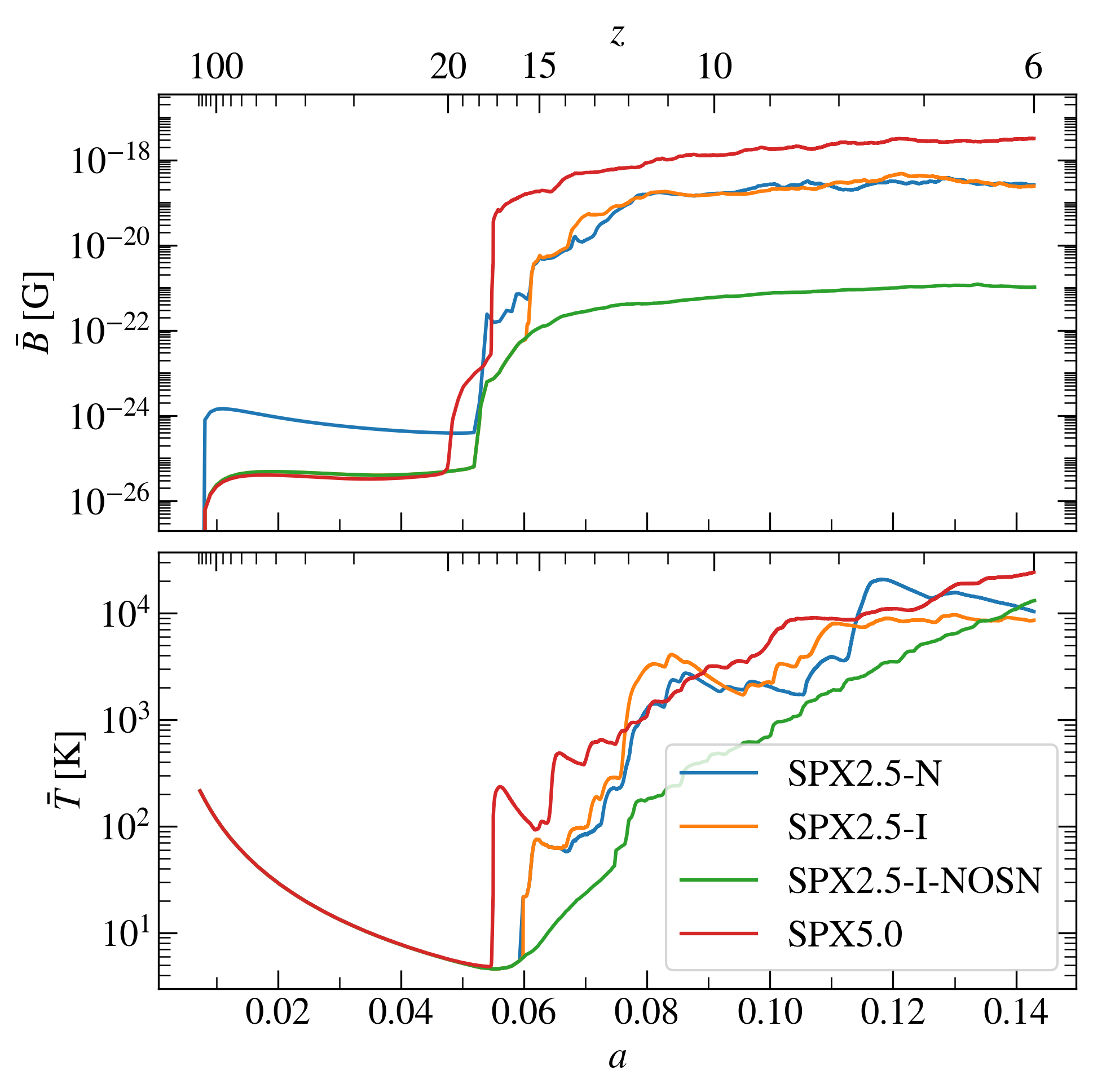}
\caption{Evolution of the volume-averaged magnetic field strength (top) and volume-averaged temperature (bottom) as a function of the scale factor (or redshift) for the different cosmological simulations.}
\label{fig:sph_z}
\end{figure}

A well-known limitation of radiation hydrodynamics simulations of the EoR is their sensitivity to the finite size of the computational box \citep[e.g.][]{Iliev2006}. The presence (or absence) of large and rare haloes can significantly affect the reionization history of the simulated Universe. To understand the sensitivity of the Biermann battery to these finite volume effects, we have performed the \textsc{spx5.0} simulation, whose volume is 8 times larger than our fiducial \textsc{spx2.5} simulations. At $z=6$, the 5 cMpc box is more efficiently ionized than the 2.5 cMpc box (volume-averaged ionized hydrogen fraction of $x_\textsc{H ii}=0.56$ for the former versus $x_\textsc{H ii}=0.19$ for the latter). The reionization of the 5 cMpc volume is significantly slower than in \citet{Rosdahl2018}. This is primarily due to the updated version of the \bpass{} SED model we use, which assumes a smaller fraction of stars living in binary systems and leads to lower escape fractions of ionizing radiation from galaxies. This will be addressed in more detail in a forthcoming paper (Rosdahl et al., \textit{in prep}).

The spatial distribution of the magnetic field intensity (Fig.~\ref{fig:sph50}, top panel) qualitatively shows the same properties as the \textsc{spx2.5-i} simulation. A few neutral regions remain with a weak magnetic field corresponding to the linear regime of the Biermann battery. Most of the volume is filled with ionized gas at $T\simeq10^4$~K and corresponds to the Biermann battery driven by ionization fronts, resulting in a field strength $B \simeq 10^{-20}$~G. Hot gas bubbles are also visible, with a magnetic field strength between 100 to 1000 times higher than the mean value at that epoch. Given the larger number of galaxies in the larger box, they are much more numerous than for our fiducial run. Some of the bubbles are however significantly larger than in the smaller box. They correspond to more massive galaxies with higher star formation rates. In the right side of the map in Fig.~\ref{fig:sph50}, we can even see some of the hot bubbles percolating.  

We thus recover the three modes of magnetic field generation by the Biermann battery, as discussed in the previous section. This is clearly confirmed by the $B-n$ histogram (bottom panel of Fig.~\ref{fig:sph50}). The higher branch that we attribute to SN-driven galactic winds is however one order of magnitude higher than for the 2.5~cMpc volume. We believe that we witness the onset of a turbulent dynamo inside the largest galaxies of our simulated volume. Our increased spatial resolution allows us to start resolving a small-scale dynamo within the turbulent ISM \citep[as in][]{MartinAlvarez2018}. Indeed, the biggest haloes tend to be better resolved in the bigger volume, because of their more massive DM component, and the highly-resolved regions are of bigger spatial extent there. Even if the resolution is not sufficiently good to see a clear exponential growth of the field \citep[see][for a more detailed discussion]{Rieder2016,Rieder2017}, we start to see a weak amplification of the field within the galaxy, which is then re-distributed on larger scales by the galactic winds. These winds pollute the IGM with metals, as well as amplified magnetic fields \citep{Bertone2006,Butsky2016,Pakmor2020}. This explains why the upper branch follows a similar $B \propto n^{2/3}$ relation, but with a higher normalisation than the 2.5~cMpc box. It should be noted that a higher temperature in the larger volume could also explain the showcased magnetic field amplification (as per Eq.~\ref{eq:Bamp}), but we have ruled out this possibility by checking that the \textsc{spx5.0} $T-n$ histogram is similar to the \textsc{spx2.5-i} one (Fig.~\ref{fig:sph25_Td}).

In order to better understand the difference in the field generation and amplification processes between the two volumes, we look at the temporal evolution of the volume-weighted average magnetic field strength and the volume-weighted average temperature for our different simulations (Fig.~\ref{fig:sph_z}). This allows us to define three different epochs of field generation and amplification.

The first epoch corresponds to the linear regime of the Biermann battery. In our simulations, it ends around $z=18$ (resp. $z=20$) for the small box (resp. the large box) when reionization starts. During these dark ages, the main thermal mechanism that takes place is a competition between Compton heating by the CMB and adiabatic cooling of the quasi-neutral gas \citep{Naoz2005}. This complex process introduces scale-dependant temperature fluctuations. The resulting misaligned density and temperature perturbations cause a low-intensity Biermann battery, with an average field strength $B \simeq 10^{-25}$~G on scales of $\sim 20$ ckpc \citep{Naoz2013}. In our simulations, these scales are resolved and the thermodynamics of the gas are properly captured, which explains why we can observe this pre-EoR magnetic field generation as predicted by \cite{Naoz2013}. As it can also be noticed in \citet{Naoz2013}, the average field strength remains roughly constant during the linear regime. Indeed, this equilibrium value stems from a competition between the Biermann process and the cosmic expansion, for which $B$ would otherwise decay as $a^{-2}$. We can also see in Fig.~\ref{fig:sph_z} that the naive method creates too much spurious field to properly capture this process.

The second epoch corresponds to the onset of cosmic reionization. Our simulation without SN feedback only captures the reionization process, with its corresponding thermal and magnetic history (see Fig.~\ref{fig:sph_z}). The mean temperature in the box steadily rises as Strömgren spheres progressively percolate. The average magnetic field quickly rises after reionization starts and stabilises to an average field strength of $B \simeq 10^{-21}$~G. The third epoch corresponds to the onset of galactic winds, when the first SNe explode in the star-forming galaxies. It can be seen as a series of discrete events in the temperature evolution. The resulting average magnetic field is two orders of magnitude higher than in the no SNe case. In these two regimes, both the naive and the improved methods predict very similar average field strengths. The effect of the box size, although relatively weak (if present at all) for the average temperature evolution, is quite important for the average magnetic field strength evolution, with an increase close to one order of magnitude in strength. As explained earlier, we speculate that the large box simulation is marginally resolving a turbulent dynamo at small scales, within the turbulent ISM of the largest galaxies, amplifying the field 10 $\times$ more than the small box.

\begin{figure}
\centering
\includegraphics[width=8.5cm]{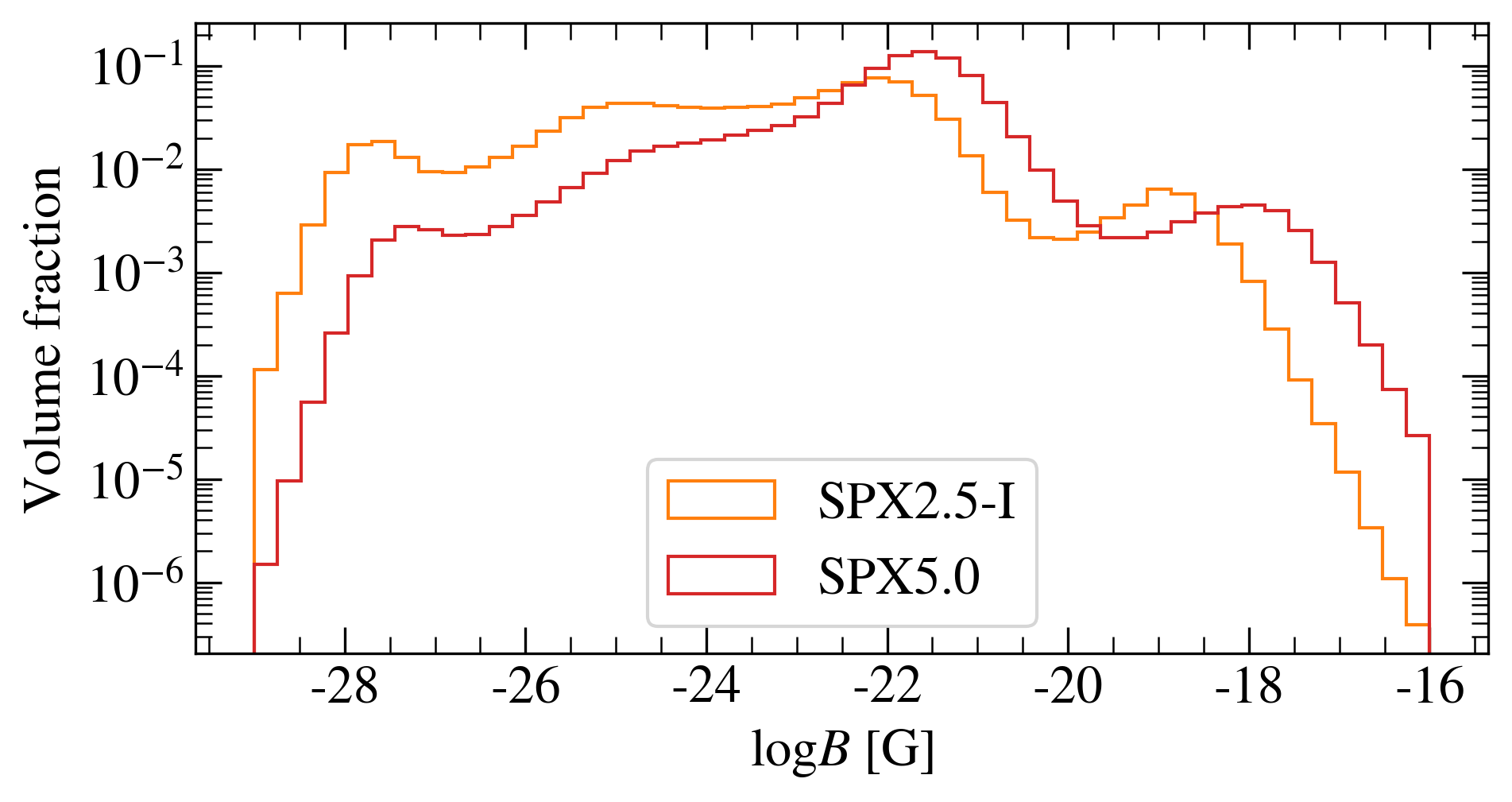}
\caption{Volume-weighted PDF of the magnetic field strength at $z=6$ for the \textsc{spx2.5-i} and \textsc{spx5.0} simulations.}
\label{fig:sph_Bvw}
\end{figure}

This conclusion is further supported by the comparison of the volume-weighted PDF of the magnetic field for the two different box sizes at $z=6$ (Fig.~\ref{fig:sph_Bvw}). First, because cosmic reionization is more efficient in the 5~cMpc box, we observe a smaller volume fraction corresponding to voids still in the linear regime with $B \le 10^{-24}$~G and a larger volume fraction corresponding to the ionization channel of the Biermann battery with $B \simeq 10^{-21}$~G. Second, both PDFs show a smaller peak at higher field strength with $B \ge 10^{-20}$~G. This secondary peak corresponds to the galactic winds with a smaller volume fraction. The small box features for this secondary peak a maximum value around $B \simeq 10^{-19}$~G, while the large box peak is clearly shifted to $B \simeq 10^{-18}$~G. This corresponds to a more efficient dynamo process in the largest galaxies and perfectly explains the average magnetic field strength temporal evolution in Fig.~\ref{fig:sph_z}. 

%%%%%%%%%%%%%%%%%%%%%%%%%%%%%%%%%%%%%%%%%%%%%%%%%%

\section{Discussion}
\label{sect:discuss}

An important aspect of this paper is the design of a robust numerical method for the Biermann battery. We have implemented and tested both a simple but naive approach and a more elaborate but accurate approach that preserves the absence of Biermann battery for isothermal flows. This structure preserving property proves to be efficient in two important test cases, namely the Strömgren sphere and the Sedov blast wave, in reducing spurious magnetic fields and accelerating the convergence of the numerical solution. In both tests, however, the presence of sharp discontinuities constitutes an important difficulty for how fast our results are converging, even for the improved method. One could certainly argue that the Biermann battery, as described by Eq.~(\ref{eq:battery}), is not a well-posed problem in the presence of such density and temperature discontinuities. We know that in fact the electron temperature $T_{\rm e}$ is actually smooth in shocks \citep{Spitzer1962,Zeldovich1967}. Indeed, because of electronic heat conduction, $T_{\rm e}$ exhibits a thermal precursor ahead of the shock that makes the temperature profile continuous. This mechanism operates on characteristic scales close to the electron mean free path, with \citep[e.g.][]{Smith2013}
\begin{equation}
\lambda_\mathrm{e} \simeq \frac{\left(k_\mathrm{B}T\right)^2}{e^4 n_\mathrm{e} \ln \Lambda},
\end{equation}
where $\ln \Lambda$ is the Coulomb logarithm. In our simulation, using for $n_{\rm e}$ the background density at $z=6$ and for $T_{\rm e}$ a temperature of $10^4$~K, (see Sect.~\ref{sect:results}), we find $\lambda_{\rm e} \simeq 70$~pc. These scales are much smaller than the resolution of our base grid (a few kpc at $z=6$), which is already quite high for such large-scale periodic box simulations. It is therefore premature to seek for the full convergence of the Biermann battery in shocks as long as $\lambda_{\rm e}$ is not resolved \citep[see a similar discussion in][]{Graziani2015}. The analysis of our test suite and our cosmological simulations suggest that the accuracy of the improved method is high enough to properly capture the Biermann battery process during the EoR.

Our cosmological simulations indeed confirm previous findings, namely that the Biermann battery is already active during the dark ages, with the linear regime of density and temperature fluctuations leading to a field strength of $B \simeq 10^{-25}$~G. We also recover the classical regime of the Biermann battery induced by the propagation of ionization fronts \citep{Subramanian1994,Gnedin2000} with $B \simeq 10^{-20}$~G. Our simulations also revealed a third magnetic field generation channel within SN-driven galactic winds. In this channel, we observe a combination of the Biermann battery within shocks and a turbulent dynamo inside galactic discs. The resulting magnetic field is between two and three orders of magnitude stronger than the field generated in ionization fronts. Note that one expects to better and better resolve the turbulent dynamo process as the resolution is increased, so that the magnetic field strength injected into the IGM by galactic winds is likely to increase until the turbulent dynamo reaches saturation within the star-forming galaxies \citep{Rieder2017,AramburoGarcia2021}. 

It is worth mentioning that the Biermann battery within shock waves has already been proposed by \cite{Kulsrud1997} and \cite{Davies2000}. In these earlier simulations, the effects are triggered by non-radiative accretion shocks around haloes and are observed at relatively low redshifts of $z < 2$. The corresponding Biermann generated field strength is only $B \simeq 10^{-20}$~G. Owing to the much stronger SN feedback recipe we used here, the shocks we observe in galactic winds are much stronger than accretion shocks, and therefore generate stronger magnetic fields. In fact, the virial temperature is on the order of $10^5$~K for the biggest haloes, as confirmed by the $T-n$ diagram (Fig.~\ref{fig:sph25_Td}). SN-driven galactic winds are typically hotter, which explains that our shock channel is stronger than classical accretion shocks.

Our results qualitatively match the recent study of \citet{Garaldi2021}, in particular the simulations they have performed with the Biermann battery. At $z=6$, the field strengths they obtain are all $\lesssim 10^{-23}$~G (see their Fig.~8), which corresponds to the linear regime of the Biermann battery. Surprisingly, no fields higher than $10^{-20}$~G appear in their simulations during the EoR but only after, once the turbulent dynamos inside galaxies saturate. The time evolution of their volume-weighted magnetic field (see their Fig.~9) is consistent with this picture, with an average field strength quite stable around $10^{-25}$~G at high redshift. We interpret this striking difference as being due to the vastly different box sizes and corresponding mass resolution between our study and theirs. It is indeed of primary importance to resolve in this context all the star-forming mini-haloes and the associated radiation and SN feedback.

We would like to stress again that we do not accurately model the turbulent dynamo process within the star-forming galaxies because we lack the required resolution \citep{Rieder2017}. At the current resolution ($\sim$ 10 pc), this would probably require a suitable sub-grid model to capture the turbulent dynamo at unresolved scales. This would result in a fast exponential amplification of the magnetic field inside galaxies, which would then be expelled and re-distributed on larger scales by galactic winds \citep{Rieder2016,MartinAlvarez2020b,Steinwandel2020}. This exponentiation would then consequently destroy all memory of the Biermann battery generated field in the regions it pollutes. However, this phenomenon is limited by the spatial extension of the outflows. Hence, it might still be possible to observe imprints of the primeval Biermann-generated fields in voids where pristine gas has not been affected by galactic winds \citep{Bertone2006,AramburoGarcia2021}.

The resulting fields from our simulations are not high enough to have an influence on gas dynamics (negligible magnetic energy in front of the total energy). In particular, they do not affect the reionization process as they have no impact on star formation for such low values \citep{Sethi2005,Marinacci2016,Katz2021}. They however provide a natural origin for seed magnetic fields subsequently amplified by the galactic dynamo. Interestingly, this process occurs before \citep{Naoz2013}, during, and after the EoR. Other seeding mechanisms have been proposed like the Durrive battery \citep{Durrive2015,Durrive2017}, for which the electric field is produced by charge segregation due to photo-ionization. It is hence powered by ionization fronts as well. The Biermann battery generates geometrically similar, but systematically stronger magnetic fields, thus being almost always the dominant process \citep{Garaldi2021}.

%%%%%%%%%%%%%%%%%%%%%%%%%%%%%%%%%%%%%%%%%%%%%%%%%%

\section{Conclusions and outlook}

We introduce in this paper a novel numerical implementation of the Biermann battery within the AMR cosmological simulation code \ramses{}. The structure-preserving nature of our scheme has the advantage of ensuring that no spurious magnetic field is generated for an isothermal flow. This property, together with the strict divergence-free nature of the CT-based ideal MHD solver in \ramses{}, proves particularly accurate for several test cases and a suite of cosmological simulations of the EoR.

Our simulation suite allowed us to study the impact of various key parameters such as the choice of the Biermann battery scheme, the presence of SN feedback and box size. In particular, we simulate two different large-scale volumes: 2.5 cMpc and 5 cMpc from $z=150$ to $z=6$. The main results can be summarised as follows.
\begin{itemize}
    \item Prior to the EoR, a small magnetic field with strength $B \simeq 10^{-25}$~G field is generated by temperature and density fluctuations induced by linear perturbations. The average field strength remains constant over time due to a competition between the Biermann battery and the expansion of the Universe. This process, predicted by \cite{Naoz2013}, is properly captured only by the improved method.
    \item During the EoR, propagating ionization fronts are able to magnetise a substantial fraction of the volume with an average field strength close to $B \simeq 10^{-20}$~G \citep[as in][]{Gnedin2000}.
    \item On top of this volume-filling field, SN-driven galactic winds around star-forming galaxies generate hot gas bubbles hosting stronger magnetic fields with $B \simeq 10^{-18}$~G around massive haloes. They are due to SN-driven shocks that are subsequently expelled into the surrounding IGM by galactic winds. Their higher strength results from the higher temperature and the more efficient Biermann battery. This channel completely disappears if SN feedback is disabled.
    \item The resulting bi-modal distribution of the magnetic field at the end of the EoR ($z=6$) is confirmed by a phase-space analysis of $B-n$ histograms with two distinct branches both scaling as $n^{2/3}$ but differently normalised with approximately a two orders of magnitude offset.
    \item The higher branch that corresponds to SN-driven galactic winds has a significantly higher normalisation in the larger box. We interpret this finite volume effect as a better resolved turbulent dynamo in the larger galaxies. The corresponding amplified field is then re-distributed on larger scales by the galactic winds.
    \item The morphology of the generated fields is complex, regularly alternating between positive and negative values. For ionization fronts as well as for galactic winds, the magnetic field tends to be mostly toroidal, circling around the regions where it is produced.
    \item The time evolution of the magnetic field is roughly consistent for all simulations. After the linear regime during the dark ages where the average field is almost constant, its time evolution reveals two major phases: the first one corresponds to the beginning of the EoR and the second one to the onset of galactic winds. Again, the second epoch is absent if SN feedback is disabled.
    \item The volume-integrated magnetic field at $z=6$ is one order of magnitude higher for the larger box. This is due to stronger fields within SN-driven bubbles, owing to a more efficient turbulent dynamo in the larger box.
\end{itemize}

The natural follow-up of the work presented here is to study in more detail the spatial distribution of the Biermann battery generated field. For this purpose, a power-spectrum analysis will be carried out in a future companion paper. This may result in a better understanding of the three different channels we have discussed here and on what spatial scales they operate. Finally, it may be interesting to couple our new numerical implementation of the Biermann battery to a sub-grid model for a small-scale dynamo, which will allow us to capture the turbulent amplification within star-forming galaxies. This will hopefully bring us one step closer to fully understanding cosmic magnetism.

%%%%%%%%%%%%%%%%%%%%%%%%%%%%%%%%%%%%%%%%%%%%%%%%%%

\section*{Acknowledgements}

We offer our thanks to our referee, Karsten Jedamzik. This work was performed with the support of the Swiss National Science Foundation (SNSF). TK was supported by the National Research Foundation of Korea (NRF-2019K2A9A1A06091377 and NRF-2020R1C1C1007079). SMA acknowledges support by the European Research Council (ERC) Starting Grant 638707 `Black holes and their host galaxies: co-evolution across cosmic time'. The simulations in this work were carried out on Piz Daint at the Swiss Supercomputing Center (CSCS) in Lugano, and the analysis was conducted with equipment maintained by the Service and Support for Science IT (S3IT), University of Zurich. We made use of the \textsc{Pynbody} package \citep{Pontzen2013}. We acknowledge the PRACE Research Infrastructure (PRACE project ID 2016153539) for providing us access to the SuperMUC computing resource based in Garching, Germany, to tune and perform the original \sphinx{} simulations on which this work expands.

%%%%%%%%%%%%%%%%%%%%%%%%%%%%%%%%%%%%%%%%%%%%%%%%%%

\section*{Data Availability}

The data underlying this article will be shared upon reasonable request to the corresponding author.



\bibliographystyle{mnras}
\bibliography{papers,romain,joki}

\begin{thebibliography}{}
\makeatletter
\relax
\def\mn@urlcharsother{\let\do\@makeother \do\$\do\&\do\#\do\^\do\_\do\%\do\~}
\def\mn@doi{\begingroup\mn@urlcharsother \@ifnextchar [ {\mn@doi@}
  {\mn@doi@[]}}
\def\mn@doi@[#1]#2{\def\@tempa{#1}\ifx\@tempa\@empty \href
  {http://dx.doi.org/#2} {doi:#2}\else \href {http://dx.doi.org/#2} {#1}\fi
  \endgroup}
\def\mn@eprint#1#2{\mn@eprint@#1:#2::\@nil}
\def\mn@eprint@arXiv#1{\href {http://arxiv.org/abs/#1} {{\tt arXiv:#1}}}
\def\mn@eprint@dblp#1{\href {http://dblp.uni-trier.de/rec/bibtex/#1.xml}
  {dblp:#1}}
\def\mn@eprint@#1:#2:#3:#4\@nil{\def\@tempa {#1}\def\@tempb {#2}\def\@tempc
  {#3}\ifx \@tempc \@empty \let \@tempc \@tempb \let \@tempb \@tempa \fi \ifx
  \@tempb \@empty \def\@tempb {arXiv}\fi \@ifundefined
  {mn@eprint@\@tempb}{\@tempb:\@tempc}{\expandafter \expandafter \csname
  mn@eprint@\@tempb\endcsname \expandafter{\@tempc}}}

\bibitem[\protect\citeauthoryear{{Aramburo Garcia}, {Bondarenko}, {Boyarsky},
  {Nelson}, {Pillepich}  \& {Sokolenko}}{{Aramburo Garcia}
  et~al.}{2021}]{AramburoGarcia2021}
{Aramburo Garcia} A.,  {Bondarenko} K.,  {Boyarsky} A.,  {Nelson} D.,
  {Pillepich} A.,   {Sokolenko} A.,  2021, arXiv e-prints, \href
  {https://ui.adsabs.harvard.edu/abs/2021arXiv210107207A} {p. arXiv:2101.07207}

\bibitem[\protect\citeauthoryear{{Bertone}, {Vogt}  \& {En{\ss}lin}}{{Bertone}
  et~al.}{2006}]{Bertone2006}
{Bertone} S.,  {Vogt} C.,   {En{\ss}lin} T.,  2006, \mn@doi [\mnras]
  {10.1111/j.1365-2966.2006.10474.x}, \href
  {https://ui.adsabs.harvard.edu/abs/2006MNRAS.370..319B} {370, 319}

\bibitem[\protect\citeauthoryear{{Biermann}}{{Biermann}}{1950}]{Biermann1950}
{Biermann} L.,  1950, Zeitschrift Naturforschung Teil A, \href
  {https://ui.adsabs.harvard.edu/abs/1950ZNatA...5...65B} {5, 65}

\bibitem[\protect\citeauthoryear{{Bisnovatyi-Kogan}, {Ruzmaikin}  \&
  {Syunyaev}}{{Bisnovatyi-Kogan} et~al.}{1973}]{1973SvA....17..137B}
{Bisnovatyi-Kogan} G.~S.,  {Ruzmaikin} A.~A.,   {Syunyaev} R.~A.,  1973,
  \sovast, \href {https://ui.adsabs.harvard.edu/abs/1973SvA....17..137B} {17,
  137}

\bibitem[\protect\citeauthoryear{{Brandenburg} \& {Subramanian}}{{Brandenburg}
  \& {Subramanian}}{2005}]{2005PhR...417....1B}
{Brandenburg} A.,  {Subramanian} K.,  2005, \mn@doi [\physrep]
  {10.1016/j.physrep.2005.06.005}, \href
  {https://ui.adsabs.harvard.edu/abs/2005PhR...417....1B} {417, 1}

\bibitem[\protect\citeauthoryear{{Butsky} et~al.,}{{Butsky}
  et~al.}{2016}]{Butsky2016}
{Butsky} I.,  et~al., 2016, \mn@doi [\mnras] {10.1093/mnras/stw1688}, \href
  {https://ui.adsabs.harvard.edu/abs/2016MNRAS.462..663B} {462, 663}

\bibitem[\protect\citeauthoryear{{Davies} \& {Widrow}}{{Davies} \&
  {Widrow}}{2000}]{Davies2000}
{Davies} G.,  {Widrow} L.~M.,  2000, \mn@doi [ApJ] {10.1086/309358}, \href
  {https://ui.adsabs.harvard.edu/abs/2000ApJ...540..755D} {540, 755}

\bibitem[\protect\citeauthoryear{{Deparis}, {Aubert}, {Ocvirk}, {Chardin}  \&
  {Lewis}}{{Deparis} et~al.}{2019}]{Deparis2019}
{Deparis} N.,  {Aubert} D.,  {Ocvirk} P.,  {Chardin} J.,   {Lewis} J.,  2019,
  \mn@doi [\aap] {10.1051/0004-6361/201832889}, \href
  {https://ui.adsabs.harvard.edu/abs/2019A&A...622A.142D} {622, A142}

\bibitem[\protect\citeauthoryear{{Dubois} \& {Teyssier}}{{Dubois} \&
  {Teyssier}}{2008}]{Dubois2008}
{Dubois} Y.,  {Teyssier} R.,  2008, \mn@doi [\aap]
  {10.1051/0004-6361:200809513}, \href
  {https://ui.adsabs.harvard.edu/abs/2008A&A...482L..13D} {482, L13}

\bibitem[\protect\citeauthoryear{{Durrer} \& {Neronov}}{{Durrer} \&
  {Neronov}}{2013}]{2013A&ARv..21...62D}
{Durrer} R.,  {Neronov} A.,  2013, \mn@doi [\aapr] {10.1007/s00159-013-0062-7},
  \href {https://ui.adsabs.harvard.edu/abs/2013A&ARv..21...62D} {21, 62}

\bibitem[\protect\citeauthoryear{{Durrive} \& {Langer}}{{Durrive} \&
  {Langer}}{2015}]{Durrive2015}
{Durrive} J.~B.,  {Langer} M.,  2015, \mn@doi [\mnras] {10.1093/mnras/stv1578},
  \href {https://ui.adsabs.harvard.edu/abs/2015MNRAS.453..345D} {453, 345}

\bibitem[\protect\citeauthoryear{{Durrive}, {Tashiro}, {Langer}  \&
  {Sugiyama}}{{Durrive} et~al.}{2017}]{Durrive2017}
{Durrive} J.-B.,  {Tashiro} H.,  {Langer} M.,   {Sugiyama} N.,  2017, \mn@doi
  [\mnras] {10.1093/mnras/stx2007}, \href
  {https://ui.adsabs.harvard.edu/abs/2017MNRAS.472.1649D} {472, 1649}

\bibitem[\protect\citeauthoryear{Eldridge, Izzard  \& Tout}{Eldridge
  et~al.}{2007}]{Eldridge2007}
Eldridge J.~J.,  Izzard R.~G.,   Tout C.~A.,  2007, \mn@doi [MNRAS]
  {10.1111/j.1365-2966.2007.12738.x}, 384, 1109

\bibitem[\protect\citeauthoryear{{Evans} \& {Hawley}}{{Evans} \&
  {Hawley}}{1988}]{Evans1988}
{Evans} C.~R.,  {Hawley} J.~F.,  1988, \mn@doi [ApJ] {10.1086/166684}, \href
  {https://ui.adsabs.harvard.edu/abs/1988ApJ...332..659E} {332, 659}

\bibitem[\protect\citeauthoryear{{Fan}, {Carilli}  \& {Keating}}{{Fan}
  et~al.}{2006}]{2006ARA&A..44..415F}
{Fan} X.,  {Carilli} C.~L.,   {Keating} B.,  2006, \mn@doi [\araa]
  {10.1146/annurev.astro.44.051905.092514}, \href
  {https://ui.adsabs.harvard.edu/abs/2006ARA&A..44..415F} {44, 415}

\bibitem[\protect\citeauthoryear{{Fatenejad} et~al.,}{{Fatenejad}
  et~al.}{2013}]{Fatenejad2013}
{Fatenejad} M.,  et~al., 2013, \mn@doi [High Energy Density Physics]
  {10.1016/j.hedp.2012.11.002}, \href
  {https://ui.adsabs.harvard.edu/abs/2013HEDP....9..172F} {9, 172}

\bibitem[\protect\citeauthoryear{{Ferland}, {Korista}, {Verner}, {Ferguson},
  {Kingdon}  \& {Verner}}{{Ferland} et~al.}{1998}]{Ferland1998}
{Ferland} G.~J.,  {Korista} K.~T.,  {Verner} D.~A.,  {Ferguson} J.~W.,
  {Kingdon} J.~B.,   {Verner} E.~M.,  1998, \mn@doi [\pasp] {10.1086/316190},
  \href {https://ui.adsabs.harvard.edu/abs/1998PASP..110..761F} {110, 761}

\bibitem[\protect\citeauthoryear{{Finkelstein} et~al.,}{{Finkelstein}
  et~al.}{2019}]{2019ApJ...879...36F}
{Finkelstein} S.~L.,  et~al., 2019, \mn@doi [\apj] {10.3847/1538-4357/ab1ea8},
  \href {https://ui.adsabs.harvard.edu/abs/2019ApJ...879...36F} {879, 36}

\bibitem[\protect\citeauthoryear{{Fromang}, {Hennebelle}  \&
  {Teyssier}}{{Fromang} et~al.}{2006}]{Fromang2006}
{Fromang} S.,  {Hennebelle} P.,   {Teyssier} R.,  2006, \mn@doi [\aap]
  {10.1051/0004-6361:20065371}, \href
  {https://ui.adsabs.harvard.edu/abs/2006A&A...457..371F} {457, 371}

\bibitem[\protect\citeauthoryear{{Fryxell} et~al.,}{{Fryxell}
  et~al.}{2010}]{2010HEDP....6..162F}
{Fryxell} B.,  et~al., 2010, \mn@doi [High Energy Density Physics]
  {10.1016/j.hedp.2010.01.008}, \href
  {https://ui.adsabs.harvard.edu/abs/2010HEDP....6..162F} {6, 162}

\bibitem[\protect\citeauthoryear{{Garaldi}, {Pakmor}  \& {Springel}}{{Garaldi}
  et~al.}{2021}]{Garaldi2021}
{Garaldi} E.,  {Pakmor} R.,   {Springel} V.,  2021, \mn@doi [\mnras]
  {10.1093/mnras/stab086}, \href
  {https://ui.adsabs.harvard.edu/abs/2021MNRAS.502.5726G} {502, 5726}

\bibitem[\protect\citeauthoryear{{Gnedin}, {Ferrara}  \& {Zweibel}}{{Gnedin}
  et~al.}{2000}]{Gnedin2000}
{Gnedin} N.~Y.,  {Ferrara} A.,   {Zweibel} E.~G.,  2000, \mn@doi [ApJ]
  {10.1086/309272}, \href
  {https://ui.adsabs.harvard.edu/abs/2000ApJ...539..505G} {539, 505}

\bibitem[\protect\citeauthoryear{{Graziani}, {Tzeferacos}, {Lee}, {Lamb},
  {Weide}, {Fatenejad}  \& {Miller}}{{Graziani} et~al.}{2015}]{Graziani2015}
{Graziani} C.,  {Tzeferacos} P.,  {Lee} D.,  {Lamb} D.~Q.,  {Weide} K.,
  {Fatenejad} M.,   {Miller} J.,  2015, \mn@doi [\apj]
  {10.1088/0004-637X/802/1/43}, \href
  {https://ui.adsabs.harvard.edu/abs/2015ApJ...802...43G} {802, 43}

\bibitem[\protect\citeauthoryear{{Graziani}, {Tzeferacos}, {Lee}, {Lamb},
  {Weide}, {Fatenejad}  \& {Miller}}{{Graziani} et~al.}{2016}]{Graziani2016}
{Graziani} C.,  {Tzeferacos} P.,  {Lee} D.,  {Lamb} D.~Q.,  {Weide} K.,
  {Fatenejad} M.,   {Miller} J.,  2016, in Journal of Physics Conference
  Series. p. 012018, \mn@doi{10.1088/1742-6596/719/1/012018}

\bibitem[\protect\citeauthoryear{Hahn \& Abel}{Hahn \& Abel}{2011}]{Hahn2011}
Hahn O.,  Abel T.,  2011, \mn@doi [MNRAS] {10.1111/J.1365-2966.2011.18820.X},
  415, 2101

\bibitem[\protect\citeauthoryear{{Iliev}, {Mellema}, {Pen}, {Merz}, {Shapiro}
  \& {Alvarez}}{{Iliev} et~al.}{2006}]{Iliev2006}
{Iliev} I.~T.,  {Mellema} G.,  {Pen} U.~L.,  {Merz} H.,  {Shapiro} P.~R.,
  {Alvarez} M.~A.,  2006, \mn@doi [\mnras] {10.1111/j.1365-2966.2006.10502.x},
  \href {https://ui.adsabs.harvard.edu/abs/2006MNRAS.369.1625I} {369, 1625}

\bibitem[\protect\citeauthoryear{{Katz}, {Kimm}, {Sijacki}  \&
  {Haehnelt}}{{Katz} et~al.}{2017}]{Katz2017}
{Katz} H.,  {Kimm} T.,  {Sijacki} D.,   {Haehnelt} M.~G.,  2017, \mn@doi
  [\mnras] {10.1093/mnras/stx608}, \href
  {https://ui.adsabs.harvard.edu/abs/2017MNRAS.468.4831K} {468, 4831}

\bibitem[\protect\citeauthoryear{{Katz} et~al.,}{{Katz}
  et~al.}{2021}]{Katz2021}
{Katz} H.,  et~al., 2021, arXiv e-prints, \href
  {https://ui.adsabs.harvard.edu/abs/2021arXiv210111624K} {p. arXiv:2101.11624}

\bibitem[\protect\citeauthoryear{{Kimm} \& {Cen}}{{Kimm} \&
  {Cen}}{2014}]{2014ApJ...788..121K}
{Kimm} T.,  {Cen} R.,  2014, \mn@doi [\apj] {10.1088/0004-637X/788/2/121},
  \href {https://ui.adsabs.harvard.edu/abs/2014ApJ...788..121K} {788, 121}

\bibitem[\protect\citeauthoryear{{Kimm}, {Katz}, {Haehnelt}, {Rosdahl},
  {Devriendt}  \& {Slyz}}{{Kimm} et~al.}{2017}]{Kimm2017}
{Kimm} T.,  {Katz} H.,  {Haehnelt} M.,  {Rosdahl} J.,  {Devriendt} J.,   {Slyz}
  A.,  2017, \mn@doi [\mnras] {10.1093/mnras/stx052}, \href
  {https://ui.adsabs.harvard.edu/abs/2017MNRAS.466.4826K} {466, 4826}

\bibitem[\protect\citeauthoryear{{Kroupa}}{{Kroupa}}{2001}]{Kroupa2001}
{Kroupa} P.,  2001, \mn@doi [\mnras] {10.1046/j.1365-8711.2001.04022.x}, \href
  {https://ui.adsabs.harvard.edu/abs/2001MNRAS.322..231K} {322, 231}

\bibitem[\protect\citeauthoryear{{Kulkarni}, {Keating}, {Haehnelt}, {Bosman},
  {Puchwein}, {Chardin}  \& {Aubert}}{{Kulkarni} et~al.}{2019}]{Kulkarni2019}
{Kulkarni} G.,  {Keating} L.~C.,  {Haehnelt} M.~G.,  {Bosman} S. E.~I.,
  {Puchwein} E.,  {Chardin} J.,   {Aubert} D.,  2019, \mn@doi [\mnras]
  {10.1093/mnrasl/slz025}, \href
  {https://ui.adsabs.harvard.edu/abs/2019MNRAS.485L..24K} {485, L24}

\bibitem[\protect\citeauthoryear{{Kulsrud}, {Cen}, {Ostriker}  \&
  {Ryu}}{{Kulsrud} et~al.}{1997}]{Kulsrud1997}
{Kulsrud} R.~M.,  {Cen} R.,  {Ostriker} J.~P.,   {Ryu} D.,  1997, \mn@doi [ApJ]
  {10.1086/303987}, \href
  {https://ui.adsabs.harvard.edu/abs/1997ApJ...480..481K} {480, 481}

\bibitem[\protect\citeauthoryear{{Lazar}, {Smolyakov}, {Schlickeiser}  \&
  {Shukla}}{{Lazar} et~al.}{2009}]{2009JPlPh..75...19L}
{Lazar} M.,  {Smolyakov} A.,  {Schlickeiser} R.,   {Shukla} P.~K.,  2009,
  \mn@doi [Journal of Plasma Physics] {10.1017/S0022377807007015}, \href
  {https://ui.adsabs.harvard.edu/abs/2009JPlPh..75...19L} {75, 19}

\bibitem[\protect\citeauthoryear{{Marinacci} \& {Vogelsberger}}{{Marinacci} \&
  {Vogelsberger}}{2016}]{Marinacci2016}
{Marinacci} F.,  {Vogelsberger} M.,  2016, \mn@doi [\mnras]
  {10.1093/mnrasl/slv176}, \href
  {https://ui.adsabs.harvard.edu/abs/2016MNRAS.456L..69M} {456, L69}

\bibitem[\protect\citeauthoryear{{Martel} \& {Shapiro}}{{Martel} \&
  {Shapiro}}{1998}]{Martel1998}
{Martel} H.,  {Shapiro} P.~R.,  1998, \mn@doi [\mnras]
  {10.1046/j.1365-8711.1998.01497.x}, \href
  {https://ui.adsabs.harvard.edu/abs/1998MNRAS.297..467M} {297, 467}

\bibitem[\protect\citeauthoryear{{Martin-Alvarez}, {Devriendt}, {Slyz}  \&
  {Teyssier}}{{Martin-Alvarez} et~al.}{2018}]{MartinAlvarez2018}
{Martin-Alvarez} S.,  {Devriendt} J.,  {Slyz} A.,   {Teyssier} R.,  2018,
  \mn@doi [\mnras] {10.1093/mnras/sty1623}, \href
  {https://ui.adsabs.harvard.edu/abs/2018MNRAS.479.3343M} {479, 3343}

\bibitem[\protect\citeauthoryear{{Martin-Alvarez}, {Katz}, {Sijacki},
  {Devriendt}  \& {Slyz}}{{Martin-Alvarez} et~al.}{2020a}]{MartinAlvarez2020b}
{Martin-Alvarez} S.,  {Katz} H.,  {Sijacki} D.,  {Devriendt} J.,   {Slyz} A.,
  2020a, arXiv e-prints, \href
  {https://ui.adsabs.harvard.edu/abs/2020arXiv201111648M} {p. arXiv:2011.11648}

\bibitem[\protect\citeauthoryear{{Martin-Alvarez}, {Slyz}, {Devriendt}  \&
  {G{\'o}mez-Guijarro}}{{Martin-Alvarez} et~al.}{2020b}]{MartinAlvarez2020a}
{Martin-Alvarez} S.,  {Slyz} A.,  {Devriendt} J.,   {G{\'o}mez-Guijarro} C.,
  2020b, \mn@doi [\mnras] {10.1093/mnras/staa1438}, \href
  {https://ui.adsabs.harvard.edu/abs/2020MNRAS.495.4475M} {495, 4475}

\bibitem[\protect\citeauthoryear{{Naab} \& {Ostriker}}{{Naab} \&
  {Ostriker}}{2017}]{2017ARA&A..55...59N}
{Naab} T.,  {Ostriker} J.~P.,  2017, \mn@doi [\araa]
  {10.1146/annurev-astro-081913-040019}, \href
  {https://ui.adsabs.harvard.edu/abs/2017ARA&A..55...59N} {55, 59}

\bibitem[\protect\citeauthoryear{{Naoz} \& {Barkana}}{{Naoz} \&
  {Barkana}}{2005}]{Naoz2005}
{Naoz} S.,  {Barkana} R.,  2005, \mn@doi [\mnras]
  {10.1111/j.1365-2966.2005.09385.x}, \href
  {https://ui.adsabs.harvard.edu/abs/2005MNRAS.362.1047N} {362, 1047}

\bibitem[\protect\citeauthoryear{{Naoz} \& {Narayan}}{{Naoz} \&
  {Narayan}}{2013}]{Naoz2013}
{Naoz} S.,  {Narayan} R.,  2013, \mn@doi [\prl]
  {10.1103/PhysRevLett.111.051303}, \href
  {https://ui.adsabs.harvard.edu/abs/2013PhRvL.111e1303N} {111, 051303}

\bibitem[\protect\citeauthoryear{{Ocvirk}, {Aubert}, {Chardin}, {Deparis}  \&
  {Lewis}}{{Ocvirk} et~al.}{2019}]{Ocvirk2019}
{Ocvirk} P.,  {Aubert} D.,  {Chardin} J.,  {Deparis} N.,   {Lewis} J.,  2019,
  \mn@doi [\aap] {10.1051/0004-6361/201832923}, \href
  {https://ui.adsabs.harvard.edu/abs/2019A&A...626A..77O} {626, A77}

\bibitem[\protect\citeauthoryear{{Pakmor} et~al.,}{{Pakmor}
  et~al.}{2020}]{Pakmor2020}
{Pakmor} R.,  et~al., 2020, \mn@doi [\mnras] {10.1093/mnras/staa2530}, \href
  {https://ui.adsabs.harvard.edu/abs/2020MNRAS.498.3125P} {498, 3125}

\bibitem[\protect\citeauthoryear{{Planck Collaboration}}{{Planck
  Collaboration}}{2014}]{Ade2014}
{Planck Collaboration} 2014, \mn@doi [A{\&}A] {10.1051/0004-6361/201321529},
  571, A1

\bibitem[\protect\citeauthoryear{{Pontzen}, {Ro{\v{s}}kar}, {Stinson}  \&
  {Woods}}{{Pontzen} et~al.}{2013}]{Pontzen2013}
{Pontzen} A.,  {Ro{\v{s}}kar} R.,  {Stinson} G.,   {Woods} R.,  2013, {pynbody:
  N-Body/SPH analysis for python} (\mn@eprint {ascl} {1305.002})

\bibitem[\protect\citeauthoryear{{Quilis}, {Mart{\'\i}}  \&
  {Planelles}}{{Quilis} et~al.}{2020}]{Quilis2020}
{Quilis} V.,  {Mart{\'\i}} J.-M.,   {Planelles} S.,  2020, \mn@doi [\mnras]
  {10.1093/mnras/staa877}, \href
  {https://ui.adsabs.harvard.edu/abs/2020MNRAS.494.2706Q} {494, 2706}

\bibitem[\protect\citeauthoryear{{Rees}}{{Rees}}{2005}]{2005LNP...664....1R}
{Rees} M.~J.,  2005, {Magnetic Fields in the Early Universe}.
{Wielebinski}, Richard and {Beck}, Rainer, \mn@doi{10.1007/11369875_1}

\bibitem[\protect\citeauthoryear{{Rieder} \& {Teyssier}}{{Rieder} \&
  {Teyssier}}{2016}]{Rieder2016}
{Rieder} M.,  {Teyssier} R.,  2016, \mn@doi [\mnras] {10.1093/mnras/stv2985},
  \href {https://ui.adsabs.harvard.edu/abs/2016MNRAS.457.1722R} {457, 1722}

\bibitem[\protect\citeauthoryear{{Rieder} \& {Teyssier}}{{Rieder} \&
  {Teyssier}}{2017}]{Rieder2017}
{Rieder} M.,  {Teyssier} R.,  2017, \mn@doi [\mnras] {10.1093/mnras/stx2276},
  \href {https://ui.adsabs.harvard.edu/abs/2017MNRAS.472.4368R} {472, 4368}

\bibitem[\protect\citeauthoryear{{Rosdahl}, {Blaizot}, {Aubert}, {Stranex}  \&
  {Teyssier}}{{Rosdahl} et~al.}{2013}]{Rosdahl2013}
{Rosdahl} J.,  {Blaizot} J.,  {Aubert} D.,  {Stranex} T.,   {Teyssier} R.,
  2013, \mn@doi [\mnras] {10.1093/mnras/stt1722}, \href
  {https://ui.adsabs.harvard.edu/abs/2013MNRAS.436.2188R} {436, 2188}

\bibitem[\protect\citeauthoryear{{Rosdahl} et~al.,}{{Rosdahl}
  et~al.}{2018}]{Rosdahl2018}
{Rosdahl} J.,  et~al., 2018, \mn@doi [\mnras] {10.1093/mnras/sty1655}, \href
  {https://ui.adsabs.harvard.edu/abs/2018MNRAS.479..994R} {479, 994}

\bibitem[\protect\citeauthoryear{{Rosen} \& {Bregman}}{{Rosen} \&
  {Bregman}}{1995}]{Rosen1995}
{Rosen} A.,  {Bregman} J.~N.,  1995, \mn@doi [\apj] {10.1086/175303}, \href
  {https://ui.adsabs.harvard.edu/abs/1995ApJ...440..634R} {440, 634}

\bibitem[\protect\citeauthoryear{{Sethi} \& {Subramanian}}{{Sethi} \&
  {Subramanian}}{2005}]{Sethi2005}
{Sethi} S.~K.,  {Subramanian} K.,  2005, \mn@doi [\mnras]
  {10.1111/j.1365-2966.2004.08520.x}, \href
  {https://ui.adsabs.harvard.edu/abs/2005MNRAS.356..778S} {356, 778}

\bibitem[\protect\citeauthoryear{{Smith}, {O'Shea}, {Voit}, {Ventimiglia}  \&
  {Skillman}}{{Smith} et~al.}{2013}]{Smith2013}
{Smith} B.,  {O'Shea} B.~W.,  {Voit} G.~M.,  {Ventimiglia} D.,   {Skillman}
  S.~W.,  2013, \mn@doi [\apj] {10.1088/0004-637X/778/2/152}, \href
  {https://ui.adsabs.harvard.edu/abs/2013ApJ...778..152S} {778, 152}

\bibitem[\protect\citeauthoryear{{Spitzer}}{{Spitzer}}{1962}]{Spitzer1962}
{Spitzer} L.,  1962, {Physics of Fully Ionized Gases}.
New York: Interscience Publishers

\bibitem[\protect\citeauthoryear{{Stanway} \& {Eldridge}}{{Stanway} \&
  {Eldridge}}{2018}]{Stanway2018}
{Stanway} E.~R.,  {Eldridge} J.~J.,  2018, \mn@doi [\mnras]
  {10.1093/mnras/sty1353}, \href
  {https://ui.adsabs.harvard.edu/abs/2018MNRAS.479...75S} {479, 75}

\bibitem[\protect\citeauthoryear{{Steinwandel}, {Dolag}, {Lesch}, {Moster},
  {Burkert}  \& {Prieto}}{{Steinwandel} et~al.}{2020}]{Steinwandel2020}
{Steinwandel} U.~P.,  {Dolag} K.,  {Lesch} H.,  {Moster} B.~P.,  {Burkert} A.,
   {Prieto} A.,  2020, \mn@doi [\mnras] {10.1093/mnras/staa817}, \href
  {https://ui.adsabs.harvard.edu/abs/2020MNRAS.494.4393S} {494, 4393}

\bibitem[\protect\citeauthoryear{{Subramanian}, {Narasimha}  \&
  {Chitre}}{{Subramanian} et~al.}{1994}]{Subramanian1994}
{Subramanian} K.,  {Narasimha} D.,   {Chitre} S.~M.,  1994, \mn@doi [MNRAS]
  {10.1093/mnras/271.1.L15}, \href
  {https://ui.adsabs.harvard.edu/abs/1994MNRAS.271L..15S} {271, L15}

\bibitem[\protect\citeauthoryear{{Teyssier}}{{Teyssier}}{2002}]{Teyssier2002}
{Teyssier} R.,  2002, \mn@doi [\aap] {10.1051/0004-6361:20011817}, \href
  {https://ui.adsabs.harvard.edu/abs/2002A&A...385..337T} {385, 337}

\bibitem[\protect\citeauthoryear{{Teyssier}, {Fromang}  \& {Dormy}}{{Teyssier}
  et~al.}{2006}]{Teyssier2006}
{Teyssier} R.,  {Fromang} S.,   {Dormy} E.,  2006, \mn@doi [Journal of
  Computational Physics] {10.1016/j.jcp.2006.01.042}, \href
  {https://ui.adsabs.harvard.edu/abs/2006JCoPh.218...44T} {218, 44}

\bibitem[\protect\citeauthoryear{Toro}{Toro}{2013}]{Toro2013}
Toro E.~F.,  2013, Riemann solvers and numerical methods for fluid dynamics: a
  practical introduction.
Springer Science \& Business Media

\bibitem[\protect\citeauthoryear{{T{\'o}th} et~al.,}{{T{\'o}th}
  et~al.}{2012}]{Toth2012}
{T{\'o}th} G.,  et~al., 2012, \mn@doi [Journal of Computational Physics]
  {10.1016/j.jcp.2011.02.006}, \href
  {https://ui.adsabs.harvard.edu/abs/2012JCoPh.231..870T} {231, 870}

\bibitem[\protect\citeauthoryear{{Trebitsch}, {Blaizot}, {Rosdahl}, {Devriendt}
   \& {Slyz}}{{Trebitsch} et~al.}{2017}]{Trebitsch2017}
{Trebitsch} M.,  {Blaizot} J.,  {Rosdahl} J.,  {Devriendt} J.,   {Slyz} A.,
  2017, \mn@doi [\mnras] {10.1093/mnras/stx1060}, \href
  {https://ui.adsabs.harvard.edu/abs/2017MNRAS.470..224T} {470, 224}

\bibitem[\protect\citeauthoryear{{Tweed}, {Devriendt}, {Blaizot}, {Colombi}  \&
  {Slyz}}{{Tweed} et~al.}{2009}]{Tweed2009}
{Tweed} D.,  {Devriendt} J.,  {Blaizot} J.,  {Colombi} S.,   {Slyz} A.,  2009,
  \mn@doi [\aap] {10.1051/0004-6361/200911787}, \href
  {https://ui.adsabs.harvard.edu/abs/2009A&A...506..647T} {506, 647}

\bibitem[\protect\citeauthoryear{\VAN{Leer}{Van}{van}~Leer}{\VAN{Leer}{Van}{van}~Leer}{1997}]{vanLeer1997}
\VAN{Leer}{Van}{van}~Leer B.,  1997, \mn@doi [Journal of Computational Physics]
  {10.1006/jcph.1997.5704}, \href
  {https://ui.adsabs.harvard.edu/abs/1997JCoPh.135..229V} {135, 229}

\bibitem[\protect\citeauthoryear{{Wise}, {Turk}, {Norman}  \& {Abel}}{{Wise}
  et~al.}{2012}]{2012ApJ...745...50W}
{Wise} J.~H.,  {Turk} M.~J.,  {Norman} M.~L.,   {Abel} T.,  2012, \mn@doi
  [\apj] {10.1088/0004-637X/745/1/50}, \href
  {https://ui.adsabs.harvard.edu/abs/2012ApJ...745...50W} {745, 50}

\bibitem[\protect\citeauthoryear{{Yee}}{{Yee}}{1966}]{Yee1966}
{Yee} K.,  1966, \mn@doi [IEEE Transactions on Antennas and Propagation]
  {10.1109/TAP.1966.1138693}, \href
  {https://ui.adsabs.harvard.edu/abs/1966ITAP...14..302Y} {14, 302}

\bibitem[\protect\citeauthoryear{{Zel'dovich} \& {Raizer}}{{Zel'dovich} \&
  {Raizer}}{1967}]{Zeldovich1967}
{Zel'dovich} Y.~B.,  {Raizer} Y.~P.,  1967, {Physics of shock waves and
  high-temperature hydrodynamic phenomena}.
Dover

\makeatother
\end{thebibliography}

%%%%%%%%%%%%%%%%%%%%%%%%%%%%%%%%%%%%%%%%%%%%%%%%%%

%%%%%%%%%%%%%%%%% APPENDICES %%%%%%%%%%%%%%%%%%%%%

%\appendix

%%%%%%%%%%%%%%%%%%%%%%%%%%%%%%%%%%%%%%%%%%%%%%%%%%

% Don't change these lines
\bsp	% typesetting comment
\label{lastpage}
\end{document}